\RequirePackage{fix-cm}
\documentclass[smallextended]{svjour3}       
\smartqed  

\usepackage[utf8]{inputenc}
\usepackage{graphicx}
\newenvironment{biography}[1]{%
  \begin{minipage}{\textwidth}%
    \includegraphics[width=1in, height=1.25in, clip, keepaspectratio]{#1}%
    \par%
}{%
  \end{minipage}%
}
\usepackage{wrapfig}
\usepackage{multirow}
\usepackage{tabularx}
\usepackage{longtable}
\usepackage{booktabs}
\usepackage{setspace}
\usepackage{fontawesome}
\usepackage{color}
\usepackage{appendix}
\usepackage {natbib}
\usepackage{url}
\usepackage{enumerate}
\usepackage{rotating}
\usepackage{amsmath}
\usepackage{array}
\usepackage{tikz}

\usepackage{marginnote}

\begin{document}

\title{Ethics in AI through the Practitioner’s View: A
Grounded Theory Literature Review
}

\titlerunning{Ethics in AI through the Practitioner’s View}        

\author{Aastha Pant         \and  Rashina Hoda  \and Chakkrit Tantithamthavorn \and  Burak Turhan 
}

\institute{A. Pant \at
              Dept. of Software Systems and Cybersecurity, Monash University, Melbourne, Australia \\
              \email{aastha.pant@monash.edu}           
           \and
           R. Hoda \at
              Dept. of Software Systems and Cybersecurity, Monash University, Melbourne, Australia \\
              \email{rashina.hoda@monash.edu}
              \and
                  C. Tantithamthavorn \at
              Dept. of Software Systems and Cybersecurity, Monash University, Melbourne, Australia \\
              \email{chakkrit@monash.edu}
              \and
                 B. Turhan \at
              Faculty of Info. Tech. and Electrical Engineering, University of Oulu, Oulu, Finland \\
              \email{burak.turhan@oulu.fi}
}

\date{Received: date / Accepted: date}

\maketitle

\begin{abstract}
The term ethics is widely used, explored, and debated in the context of developing Artificial Intelligence (AI) based software systems. In recent years, numerous incidents have raised the profile of ethical issues in AI development and led to public concerns about the proliferation of AI technology in our everyday lives. But what do we know about the views and experiences of those who develop these systems -- the AI practitioners? We conducted a grounded theory literature review (GTLR) of 38 primary empirical studies that included AI practitioners' views on ethics in AI and analysed them to derive five categories: practitioner \textit{awareness}, \textit{perception}, \textit{need}, \textit{challenge}, and \textit{approach}. These are underpinned by multiple codes and concepts that we explain with evidence from the included studies. We present a \textit{taxonomy of ethics in AI from practitioners' viewpoints} to assist AI practitioners in identifying and understanding the different aspects of AI ethics. The taxonomy provides a landscape view of the key aspects that concern AI practitioners when it comes to ethics in AI. We also share an agenda for future research studies and recommendations for practitioners, managers, and organisations to help in their efforts to better consider and implement ethics in AI.
\keywords{artificial intelligence \and ethics \and AI ethics \and grounded theory literature review \and practitioners \and software engineering }
\end{abstract}

\section{Introduction} \label{Introduction}
Over the last few years, there has been a swift rise in the adoption of AI technology across diverse sectors such as health, transportation, education, IT, banking, and more. The widespread use of AI has underscored the significance of ethical considerations within the realm of AI \citep{hagendorff2020ethics}. Ethics refers to ``\textit{the moral principles that govern the behaviors or activities of a person or a group of people}'' \citep{nalini2020hitchhiker}. The process of attributing moral values and ethical principles to machines to resolve ethical issues they encounter, and enabling them to operate ethically is a form of applied ethics \citep{anderson2011machine}. There is a lack of a universal definition of AI ethics and ethical principles \citep{kazim2021high}. In our study, we adopted the definition proposed by \citet{siau2020artificial}, stating that \textit{``AI ethics refers to the principles of developing AI to interact with other AIs and humans ethically and function ethically in society"}. Likewise, we have adopted the definitions of AI ethical principles outlined in Australia’s AI Ethics Principles\footnotemark{} list because there is a lack of a universal set of AI ethics principles that the whole world follows. Different countries and organisations have their own distinct AI ethical principles. For example, the European Commission has defined its own guidelines for trustworthy AI \citep{EUethics}, the United States Department of Defense has adopted 5 principles of AI Ethics \citep{USethics}, and the Organisation for Economic Cooperation and Development (OECD) has defined its AI principles to promote the use of ethical AI \citep{OECDethics}. Australia’s AI Ethics Principles address a broad spectrum of ethical concerns, spanning from human to environmental well-being. They encompass widely recognised ethical principles like fairness, privacy, and transparency, along with less common but crucial concepts such as contestability and accountability. The definitions of the terminologies used in this study have been provided in Appendix \ref{Appendix C}. \footnotetext{https://www.industry.gov.au/publications/australias-artificial-intelligence-ethics-framework/australias-ai-ethics-principles}

The consideration of ethics in AI includes the process of development as well as the resulting product\footnote{Throughout the manuscript we use the term “product” for simplicity to refer to both “products and services” where the distinction is usually straightforward from context. Also, the term `AI development' encompasses both the development and implementation of new and existing AI methods and the use of AI methods as a key component as part of a broader system.}. It is very important to incorporate ethical considerations in the development of AI products to ensure that the end product is ethically, socially, and legally responsible \citep{obermeyer2016predicting}. The importance of ethical consideration in AI is highlighted by recent incidents that demonstrate its impact \citep{bostrom2018ethics}. For example, \textit{GitHub} was criticised for using unlicensed source code as training data for their AI product, which resulted in disappointment among software developers \citep{al2023ab}. There were also cases of racial and gender bias in AI systems, such as facial recognition algorithms that performed better on white men and worse on black women, highlighting issues of accountability and bias \citep{buolamwini2018gender}. Additionally, in 2018, \textit{Amazon} had to halt the use of their AI-powered recruitment tool due to gender bias \citep{Amazon}, and in 2020, the Dutch court halted the use of System Risk Indication (SyRI) - a secret algorithm to detect possible social welfare fraud as this algorithm lacked transparency for citizens about what it does with the personal information of the people \citep{SyRI}. In each of these examples, ethical problems might have arisen during the development process, giving rise to ethical concerns regarding the resulting product. These incidents emphasise the importance of ethical considerations in AI development.

We were motivated to study the area of ethics in AI due to various case studies and the importance of the topic. Despite the existence of ethical principles, guidelines, and company policies, the implementation of these principles is ultimately up to the AI practitioners. Thus, we became interested in conducting a review study to explore existing research on ethics in AI. Specifically, we were interested in exploring the perspectives of those closest to it -- the AI practitioners\footnote{The term `practitioners' in our study includes AI developers, AI engineers, AI specialists, and AI experts. The terms `AI practitioners' and `practitioners' are used interchangeably throughout our study.}, as they are in a unique position to bring about changes and improvements and the need for review studies in the area of AI ethics to understand practitioners' perspectives have also been highlighted in the literature \citep{khan2021ethics, leikas2019ethical}. 

To understand practitioners' views on AI ethics as presented in the literature, we conducted a grounded theory literature review (GTLR) following the five-step framework of \textit{define}, \textit{search}, \textit{select}, \textit{analyse}, and \textit{present} proposed by \citet{wolfswinkel2013using}. We first defined the overarching research question (RQ), \textit{What do we know from the literature about the AI practitioners' views and experiences of ethics in AI?}\footnote{We chose the term `AI system’ as an overarching way of capturing both AI and ML-based systems and this is based on the fact that all these \textit{seed} papers that we included in our study are focused on either AI, ML, or both.} Our study aimed to find empirical studies that focused on capturing the views and experiences of AI practitioners regarding AI ethics and ethical principles, and their implementation in developing AI-based systems. Then, we used the grounded theory literature review (GTLR) protocol to search and select \emph{primary research articles}\footnote{This study uses the term ‘primary research articles’ to denote empirical works where AI practitioners were directly approached for their perspectives.} that include practitioners' views on AI ethics. To analyse the selected studies, we applied the procedures of socio-technical grounded theory (STGT) for data analysis \citep{hoda2021socio} such as \textit{open coding}, \textit{targeted coding}, \textit{constant comparison}, and \textit{memoing}, iteratively on the 38 primary empirical studies. \citet{wolfswinkel2013using} welcome adaptations to their framework by acknowledging that \textit{``... one size does not fit all, and there should be no hesitation whatsoever to deviate from our proposed steps, as long as such variation is well motivated."} Since there was little concrete guidance available on how to perform in-depth analysis and develop theory from literature as a data source, we made some adaptations, as explained in the methodology section (Section \ref{methodology}).

Based on our analysis, we present a \textit{taxonomy of ethics in AI from practitioners' viewpoints} spanning five categories: \textit{(i) practitioner awareness, (ii) practitioner perception, (iii) practitioner need, (iv) practitioner challenge, and (v) practitioner approach}, captured in Figures \ref{fig:taxonomy} and \ref{fig:overview}, and described in-depth in Sections \ref{findings} and \ref{taxonomy}. The main contributions of this paper are:

 \begin{itemize}
     \item A source of gathered information from literature on AI practitioners' views and experiences of ethics in AI,
     \item \textit{A taxonomy of ethics in AI from practitioners' viewpoints} which includes five categories such their \textit{awareness}, \textit{perception}, \textit{need}, \textit{challenge}, and \textit{approach} related to ethics in AI,
     \item An example of the application of grounded theory literature review (GTLR) in software engineering,
     \item Guidance for practitioners who require a better understanding of the requirements and factors affecting ethics implementation in AI,
     \item A set of recommendations for future research in the area of ethics implementation in AI from practitioners' perspective.

 \end{itemize}

The rest of the paper is structured as follows: Section \ref{background} presents the background details in the area of ethics in Information and Communications Technology (ICT), software engineering, and AI, followed by the details of the grounded theory literature review (GTLR) methodology in Section \ref{methodology}. Then, we discuss the challenges, threats, and limitations of the methodology in Section \ref{threats}, present the findings in Section \ref{findings} which is followed by the description of the \textit{taxonomy}, insights, and recommendations in Section \ref{discussion}. Then, we present the methodological lessons learned in section \ref{lessons} followed by a conclusion in Section \ref{conclusion}. 

\section{Background} \label{background}
\subsection{Ethics in ICT and Software Engineering}
The topic of `ethics' has been a well-researched and widely discussed topic in the field of ICT for a long time. Over recent years, various IT professional organisations worldwide, like the Association for Computing Machinery (ACM)\footnote{https://www.acm.org/code-of-ethics}, the Institute for Certification of IT Professionals (ICCP)\footnote{https://www.iccp.org/}, and AITP \footnote{https://aitp-ncfl.org/home/} have developed their own codes of ethics \citep{payne2006uniform}. These codes of ethics in the ICT domain are created to motivate and steer the ethical behavior of all computer professionals. This includes those who are currently working in the field, those who aspire to do so, teachers, students, influencers, and anyone who makes significant use of computer technology, as defined by the Association for Computing Machinery (ACM).

In 1991, \citet{gotterbarn1991computer} expressed concern about the insufficient emphasis placed on professional ethics in guiding the daily activities of computing professionals within their respective roles. Subsequently, he actively engaged in various initiatives aimed at advocating for ethical codes and fostering a sense of professional responsibility in the field. Studies have been conducted to explore how these codes of ethics affect the decision-making of professionals in the ICT sector. Ethics within the professional sphere can significantly aid ICT professionals in their decision-making, as evidenced by research conducted by \citet{allen2011information}, and these codes have been observed to influence the conduct of ICT professionals \citep{harrington1996effect}. In 2010, \citet{van2010ethical} conducted a survey involving 276 ICT professionals to explore the potential value of ethical codes of conduct for the ICT industry in dealing with contentious issues. They concluded that having a policy regarding ICT ethics does indeed significantly influence how professionals assess ethical or unethical situations in some cases. \citet{fleischmann2017societal} conducted a mixed-method study with ICT professionals on the role of codes of ethics and the relationship between their experiences and attitudes towards the codes of ethics. 

Likewise, studies have been conducted to investigate the impact of ethics in the area of Software Engineering. \citet{rashid2009software} concluded that ethics has been a very important part of software engineering and discussed the ethical challenges of software engineers who design systems for the digital world. \citet{aydemir2018roadmap} introduced an analytical framework to aid stakeholders including users and developers in capturing and analysing ethical requirements to foster ethical alignment within software artifacts and the development processes. In a similar vein, according to \citet{pierce1996computer}, one's personal ethical principles, workplace ethics, and adherence to formal codes of conduct all play a significant role in influencing the ethical conduct of software professionals. \citet{pierce1996computer} also delves into the extent of influence exerted by these three factors. On a related note, \citet{hall2009ethical} examines the concept of ethical conduct in the context of software engineers, emphasizing the importance of good professional ethics. Furthermore, in a study by \citet{fraga2022ethical}, they conducted a survey involving software engineering professionals to explore the role of ethics in their field. The findings of the study suggest that the promotion of ethical leadership among systems engineers can be achieved when they adhere to established standards, codes, and ethical principles. These studies into ethics within the realms of ICT and Software Engineering indicate that this subject has been of significant importance for a long time, and there has been a prolonged effort to improve ethical considerations in these fields.

In summary, there is a recognised need for a stronger focus on professional ethics in guiding the daily activities of computing professionals. Multiple studies consistently demonstrate the substantial influence of ethical codes on decision-making in the ICT sector and Software Engineering, shaping behavior and ethical assessments. The collective findings underscore the importance of ethical considerations in the fields of ICT and Software Engineering. 

\subsection{Secondary studies on AI Ethics}
A number of secondary studies have been conducted that focused on the theme of investigating the ethical principles and guidelines related to AI. For example, \citet{khan2021ethics} conducted a Systematic Literature Review (SLR) to investigate the agreement on the significance of AI ethical principles and identify potential challenges to their adoption. They found that the most common AI ethics principles are transparency, privacy, accountability, and fairness. However, significant challenges in incorporating ethics into AI include a lack of ethical knowledge and vague principles. Likewise, \citet{ryan2020artificial} conducted a review study to provide a comprehensive analysis of the normative consequences associated with current AI ethics guidelines, specifically targeting AI developers and organisational users. \citet{lu2022towards} conducted a Systematic Literature Review (SLR) to identify the responsible AI principles discussed in the existing literature and to uncover potential solutions for responsible AI. Additionally, they outlined a research roadmap for the field of software engineering with a focus on responsible AI.

Likewise, review studies have been conducted to investigate the ethical concerns of the use of AI in different domains. \citet{mollmann2021alright} conducted a Systematic Literature Review (SLR) to explore which ethical considerations of AI are being investigated in digital health and classified the relevant literature based on the five ethical principles of AI including \emph{beneficence,non-maleficence, autonomy, justice, and explicability}. Likewise, \citet{royakkers2018societal} conducted an SLR to explore the social and ethical issues that arise due to digitization based on six different technologies like Internet of Things, robotics, bio-metrics, persuasive technology, virtual \& augmented reality, and digital platforms. The review uncovered recurring themes such as privacy, security, autonomy, justice, human dignity, control of technology, and the balance of powers. 

Studies have also been conducted to explore different methods and approaches to enhance the ethical development of AI. For example, \citet{wiese2023being} conducted a Systematic Literature Review (SLR) to explore the methods to promote and engage practice on the front end of ethical and responsible AI. The study was guided by an adaption of the PRISMA framework and Hess \& Fore’s 2017 methodological approach. \citet{morley2020initial} conducted a review study with the aim of exploring AI ethics tools, methods, and research that are accessible to the public, for translating ethical principles into practice. 

Most of the secondary studies have either focused on investigating specific AI ethical principles, the ethical consequences of AI systems, or the approaches to enhance the ethical development of AI. Conducting a review study to identify and analyse primary empirical research on AI practitioners’ perspectives regarding AI ethics is important for gaining an understanding of the ethical landscape in the field of AI. It can also inform practical interventions, contribute to policy development, and guide educational initiatives aimed at promoting responsible and ethical practices in the development and deployment of AI technologies.

\subsection{Ethics in AI}
There are numerous and divergent views on the topic of ethics in AI \citep{vakkuri2020current,mittelstadt2019principles,hagendorff2020ethics}, as it has been increasingly applied in various contexts and industries \citep{kessing2021fairness}. AI practitioners and researchers seem to have mixed perspectives about AI ethics. Some believe there is no rush to consider AI-related ethical issues as AI has a long way from being comparable to human capabilities and behaviors \citep{siau2020artificial}, while others conclude that AI systems must be developed by considering ethics as they can have enormous societal impact \citep{bostrom2018ethics,bryson2017standardizing}. Although the viewpoints vary from practitioner to practitioner, most conclude that AI ethics is an emerging and widely discussed topic and a current relevant issue of the real world \citep{vainio2020role}. This indicates that while opinions on the importance of AI ethics may differ, there is a consensus that the subject is highly relevant in the present context.

A number of studies conducted in the area of ethics in AI have been conceptual and theoretical in nature \citep{seah2021communicating}. Critically, there are copious numbers of guidelines on AI ethics, making it challenging for AI practitioners to decide which guidelines to follow. Unsurprisingly, studies have been conducted to analyse the ever-growing list of specific AI principles \citep{kelleyemployee,mark2019ethics,siau2020artificial}. For example, \citet{jobin2019global} reviewed 84 ethical AI principles and guidelines and concluded that only five AI ethical principles -- \textit{transparency}, \textit{fairness}, \textit{non-maleficence}, \textit{responsibility} and \textit{privacy} -- are mainly discussed and followed. \cite{fjeld2020principled} reviewed 36 AI ethical principles and reported that there are eight key themes of AI ethics -- \textit{privacy}, \textit{accountability}, \textit{safety and security}, \textit{transparency} and \textit{explainability}, \textit{fairness and non-discrimination}, \textit{human control of technology}, \textit{professional responsibility}, and \textit{promotion of human values}. Likewise, \cite{hagendorff2020ethics} analysed and compared 22 AI ethical guidelines to examine their implementation in the practice of research, development, and application of AI systems. Some review studies focused on exploring the challenges and potential solutions in the area of ethics in AI, for example, \cite{jameel2020ethics,khan2021ethics}. The desire to set ethical guidelines in AI has been enhanced due to increased competition between organisations to develop robust AI tools \citep{vainio2020role}. Among them, only a few guidelines indicate an oversight or enforcement mechanism \citep{Inventory}. It suggests that recent research has dedicated significant attention to the analysis and comparison of various sets of ethical principles and guidelines for AI.

Similarly, AI practitioners have expressed various concerns regarding the public policies and ethical guidelines related to AI. For example, while the \emph{ACM Codes of Ethics} puts responsibilities to AI practitioners creating AI-based systems, a research study revealed that these practitioners generally believe that only physical harm caused by AI systems is crucial and should be taken into account \citep{veale2018fairness}. Similarly, in November 2021, the UN Educational, Scientific, and Cultural Organisation (UNESCO) signed a historic agreement outlining shared values needed to ensure the development of Responsible AI \citep{UN}. The study conducted by \citet{varanasi2023currently} involved interviewing 23 AI practitioners from 10 organisations to investigate the challenges they encounter when collaborating on Responsible AI (RAI) principles defined by UNESCO. The findings revealed that practitioners felt overwhelmed by the responsibility of adhering to specific RAI principles (non-maleficence, trustworthiness, privacy, equity, transparency, and explainability), leading to an uneven distribution of their workload. Moreover, implementing certain RAI principles (accuracy, diversity, fairness, privacy, and interoperability) in real-world scenarios proved difficult due to conflicts with personal and team values. Similarly, a study by \citet{rothenberger2019relevance} conducted an empirical study with AI experts to evaluate several AI ethics guidelines among which Microsoft AI Ethical Principles were one of them. The study found that the participants considered \emph{`Responsibility'} to be the foremost and notably significant ethical principle in the realm of AI. Following closely, they ranked \emph{`Privacy protection'} as the second most crucial principle among all other principles. This emphasises the perspective of these AI experts, who consider prioritising responsible AI practices and safeguarding user privacy to be fundamental aspects of ethical advancement and implementation of AI, without regarding other principles as equally crucial. Likewise, an empirical investigation was carried out by \citet{sanderson2023ai}, involving AI practitioners and designers. This study aimed to assess the Australian Government's high-level AI principles and investigate how these ethical guidelines were understood and applied by AI practitioners and designers within their professional contexts. The results indicated that implementing certain AI ethical principles, such as those related to \emph{`Privacy and security'}, \emph{`Transparency'} and \emph{`Explainability'}, and \emph{`Accuracy'}, posed significant challenges for them. This suggests that there have been studies exploring the relationship between AI practitioners and the guidelines established by public organisations, as well as their sentiments towards each guideline.

Another prominent area of focus has been studies that were conducted to discuss the existing gap between research and practice in the field of ethics in AI. \citet{smith2020enhancing} conducted a review study to identify gaps in ethics research and practice of ethical data-driven software development and highlighted how ethics can be integrated into the development of modern software. Similarly, \cite{shneiderman2020bridging} provided 15 recommendations to bridge the gap between ethical principles of AI and practical steps for ethical governance. Likewise, there are solution-based papers and papers discussing models, frameworks, and methods for AI developers to enhance their AI ethics implementation. For example, an article by \cite{vakkuri2021time} presents the AI maturity model for AI software. In contrast, another article by \cite{vakkuri2020eccola} discusses the ECCOLA method for implementing ethically aligned AI systems. There are also papers presenting the toolkit to address fairness in ML algorithms \citep{castelnovo2020befair} and transparency model to design transparent AI systems \citep{felzmann2020towards}. In general, it suggests that recent studies have centered on addressing the gap between research and practical application in the field of AI ethics. This also involves the development of various tools and methods aimed at improving the ethical implementation of AI.

Overall, existing studies seem to primarily focus on either analysing the plethora of ethical AI principles, filling the gap between research and practice, or discussing tool-kits and methods.  
However, compared to the number of papers on AI ethics describing ethical guidelines and principles, and tools and methods, there is a relative lack of studies that focus on the views and experiences of AI practitioners on AI ethics \citep{vakkuri2020current}. Furthermore, the literature also underscores the necessity for review studies that evaluate and synthesise the existing primary research on AI practitioners' views and experiences of AI ethics \citep{khan2021ethics, leikas2019ethical}. To assimilate, analyse, and present the empirical evidence spread across the literature, we conducted a Grounded Theory Literature Review (GTLR) to investigate AI practitioners' viewpoints on ethics in AI with some adaptations to the original framework, drawing data from papers whose prime focus may not have been understanding practitioners' viewpoints but that nonetheless contained information about the same.

\begin{figure}[htbp]
    \centering
    \includegraphics[scale=0.3]{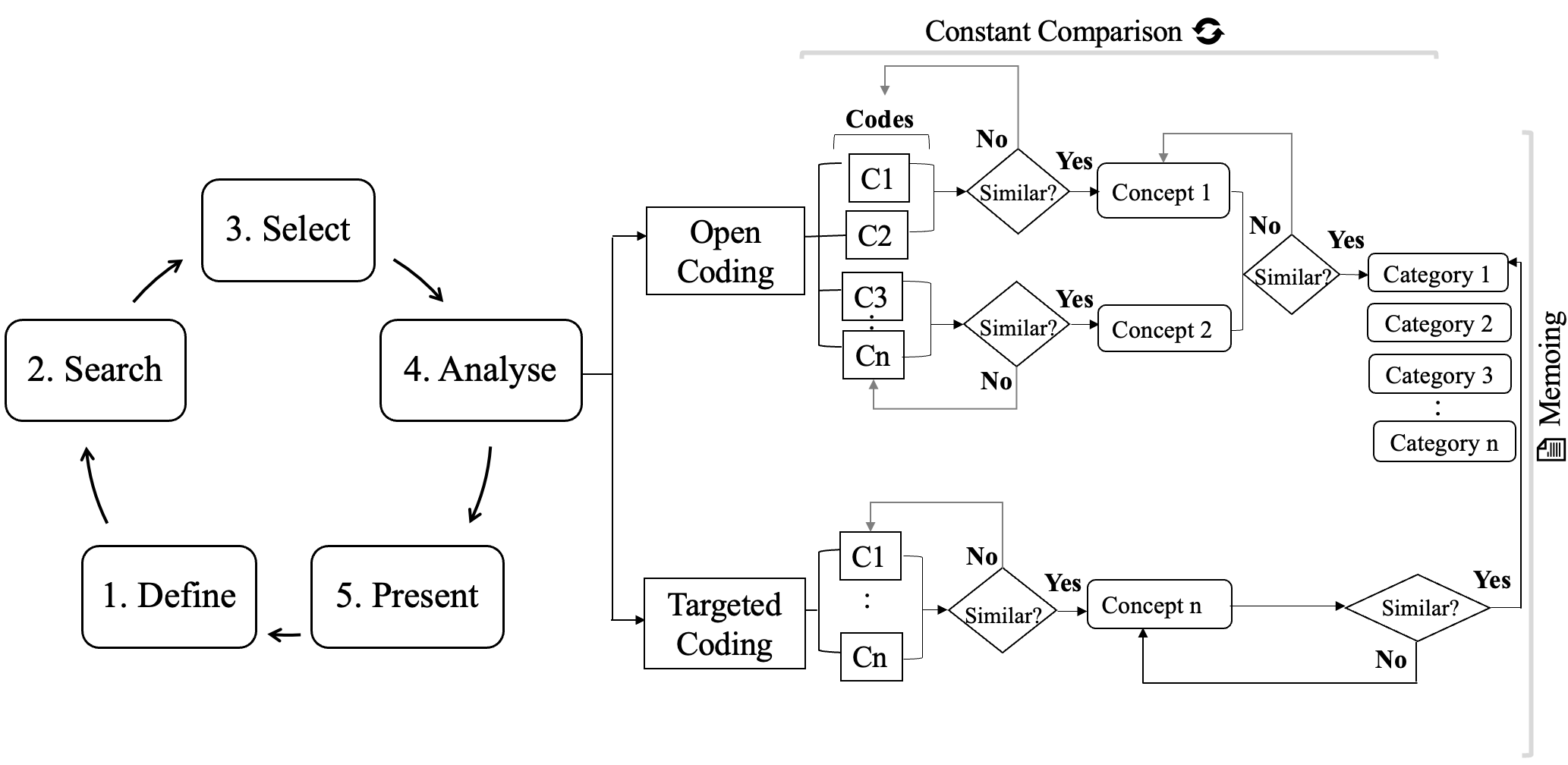}
    \caption{Steps of the Grounded Theory Literature Review (GTLR) method with Socio-Technical Grounded Theory (STGT) for data analysis}
    \label{fig:gtlr}
\end{figure}

\section{Review Methodology} \label{methodology}
While the importance of understanding AI practitioners' viewpoints on ethics in AI has been highlighted \citep{vakkuri2020current}, yet, there are not enough dedicated research articles on the topic to effectively conduct a systematic literature review or mapping study. This is mainly because there are not enough papers dedicated to investigating AI practitioners' views on ethics in AI such that their focus could be apparent from the title and abstract. Papers that include this as part of their findings are difficult to identify and select without a full read-through, making it ineffective and impractical when dealing with thousands of papers. At the same time, we were aware of a more responsive yet systematic method for reviewing the literature, called grounded theory literature review (GTLR) introduced by \cite{wolfswinkel2013using}. 
GT is a popular research method that offers a pragmatic and adaptable approach for interpreting complex social phenomena, \citep{charmaz2000grounded}. It provides a robust intellectual rationale for employing qualitative research to develop theoretical analyses \citep{goulding1998grounded}. In Grounded Theory, researchers refrain from starting with preconceived hypotheses or theories to validate or invalidate. Instead, they initiate the research process by gathering data within the context, conducting simultaneous analysis, and subsequently formulating hypotheses \citep{strauss1990basics}. This method is appropriate for our study because our research topic incorporates socio-technical aspects, and we also chose not to commence with a predetermined hypothesis. Instead, our approach was centered on examining the viewpoints of AI practitioners regarding AI ethics as outlined in the existing literature.

While the overarching review framework of grounded theory literature review  (GTLR) helped frame the review process, we found ourselves having to work through the concrete application details using the practices of socio-technical grounded theory (STGT). In doing so, we made some adaptations to the five-step framework of \textit{define}, \textit{search}, \textit{select}, \textit{analyse}, and \textit{present} described in the original grounded theory literature review (GTLR) guidelines by \cite{wolfswinkel2013using} and applied socio-technical grounded theory (STGT)'s concrete data analysis steps \citep{hoda2021socio}. Figure \ref{fig:gtlr} presents an overview of the grounded theory literature review (GTLR) steps using the socio-technical grounded theory (STGT) method for data analysis as applied in this study. Table \ref{Table 1} presents the comparison between Grounded Theory Literature Review (GTLR) as we applied it, and traditional Systematic Literature Review (SLR) \citep{kitchenham2009systematic}.

\begin{table}[ht]
\centering
\caption{Comparison of Grounded Theory Literature Review \citep{wolfswinkel2013using}, \citep{hoda2021socio} and Systematic Literature Review \citep{kitchenham2009systematic}}
\label{Table 1}
\scriptsize
\begin{tabular} {>{\raggedright\arraybackslash}p{1.4cm}>{\raggedright\arraybackslash}p{4.6cm}>{\raggedright\arraybackslash}p{4.6cm}} 

\hline

& \textbf{Systematic Literature Review (SLR)} & \textbf{Grounded Theory Literature Review (GTLR)}\\
\hline

\textbf{Definition} & Systematic review to present comprehensive findings on well-researched topics. & Rigorous review to present multi-dimensional findings and develop theoretical foundations for niche and emerging topics.\\
\hline
\textbf{Context of use} & Comprehensive coverage of well-researched topics to establish the state-of-the-art. & In-depth coverage to establish theoretical foundations and a periodic sense of the lay of the land, especially to establish an early sense of where the field is headed for niche and emerging topics.\\
\hline
\textbf{Approach} & Top-down/deductive, mostly sequential, and specification driven. & Bottom-up/inductive, iterative, and responsive.\\
\hline
\multirow{10}{*}{\textbf{Steps}} & \textit{Phase 1. Planning the review} & \textit{Iterative steps of:} \\
& \hspace{0.05cm} 1. Identify need for review & \faCaretRight \hspace{0.05cm} Define/refine RQ(s) and protocol \\
& \hspace{0.05cm} 2. Develop review protocol & \faCaretRight \hspace{0.05cm} Conduct Search\\
& \textit{Phase 2. Conducting the review} & \faCaretRight \hspace{0.05cm} Select articles  \\
& \hspace{0.05cm} 1. Identification of research & \faCaretRight \hspace{0.05cm} Conduct STGT data analysis \\
& \hspace{0.05cm} 2. Selection of primary studies & \faCaretRight \hspace{0.05cm} (optional) Develop theory or theoretical models\\
& \hspace{0.05cm} 3. Study quality assessment & \faCaretRight \hspace{0.05cm} Present findings\\
& \hspace{0.05cm} 4. Data extraction and monitoring \\
& \hspace{0.05cm} 5. Data synthesis (e.g. meta/thematic analysis) & \\
& \textit{Phase 3. Reporting the review} & \\
\hline
\textbf{Outcomes} & Full coverage findings and meta-findings. & In-depth descriptive findings, theoretical models, and theories.\\
\hline
\textbf{Effort} & Significant time and effort applied in Phases 1 and 2 (steps 1-4) & Time and effort spread across steps, more in analysis.\\ 
\hline
\textbf{Advantages} & Repeatable, reproducible, breadth of coverage, provides an overview of the aggregated information about the studies such as a number of primary studies, publication venue, year, an overview of the category of key findings etc. & Credible, rigorous, theoretical, depth of findings, can work with established and new topics, can focus on the analysis and synthesis of empirical findings in the included primary studies. \\
\hline
\textbf{Limitations} & Considerable effort involved, the possibility of biases requires established topics and can be monotonous. & Considerable effort involved, the possibility of biases, difficult to replicate outcomes.\\
\hline
\end{tabular}
\end{table}

\subsection{Define} 
The first step of grounded theory literature review (GTLR) is to formulate the initial review protocol, including determining the scope of the study by defining inclusion and exclusion criteria and search items, followed by finalising databases and search strings, with the aim of obtaining as many relevant primary empirical studies as possible. Studies that are empirical were one of the inclusion criteria of our study which is presented in Table \ref{Table 3}. By `empirical papers', we are referring to those that draw information directly from primary sources, such as interviews and survey papers (studies that involve participants by using surveys to gather their perspectives on a specific subject, not literature surveys.) The research question (RQ) formulated was, \textit{What do we know from the literature about the AI practitioners' views and experiences of ethics in AI?}

\subsubsection{Sources}
Four popular digital databases, namely, \textit{ACM Digital Library (ACM DL), IEEE Xplore, SpringerLink, and Wiley Online Library (Wiley OL)} were used as sources to identify the relevant literature. This choice was driven by the interdisciplinary nature of the topic, `ethics in AI.' Given the rapid expansion of literature on AI ethics in recent years, researchers have been contributing their work to different venues. We were interested in understanding how AI practitioners perceive AI ethics. This emphasis on AI ethics perspectives was particularly prominent within Software Engineering and Information Systems venues. These databases have also been regularly used to conduct reviews on human aspects of software engineering, for example, \cite{hidellaarachchi2021effects,perera2020study}. Initially, we searched for relevant studies which were published in journals and conferences only and for which full texts were available.

\subsubsection{Search Strings}
To begin with, we initiated the process of developing search queries by selecting key terms related to our research topic. Our initial set of key terms included ``ethics", ``AI", and ``developer". This choice was made in line with the primary objective of our study, which was to investigate the perspectives of AI practitioners on ethics in AI. Subsequently, we expanded our search by incorporating synonyms for these key terms to ensure a more comprehensive retrieval of relevant primary studies. As we constructed the final search string, we employed Boolean operators `AND' and `OR' to link these search terms. However, using the terms ``ethics", ``AI", and ``developer", along with their synonyms, resulted in a large number of papers that proved impractical to review, as illustrated in Appendix \ref{Appendix B}. In an attempt to reduce the number of papers to a manageable level, we used the term ``ethic*" along with synonyms for ``AI" and ``developer". Unfortunately, this approach yielded no results in some databases, as detailed in Appendix \ref{Appendix B}. Therefore, it became imperative for us to develop a search query that would provide us with a reasonable number of relevant primary studies to effectively conduct our study.

Six candidate search strings were developed and executed on databases before one was finalised. Table \ref{Table 2} shows the initial and final search strings. As the finalised search string returned an extremely large number of primary studies (N=9,899), we restricted the publication period from January 2010 to September 2022, in all four databases, as the topic of ethics in AI has been gaining rapid prominence in the last ten years. Table \ref{Table 3} shows the seed and final protocols, including inclusion and exclusion criteria \citep{wolfswinkel2013using}. 

\begin{table}[ht]
\caption{Formulation of search string} \label{Table 2}  
\scriptsize
\resizebox{\columnwidth}{!}{%
\begin{tabular}{l}
\hline\noalign{\smallskip}
\textbf{First search string:} (“ethics” OR “trust” OR “morals” OR “fairness” OR\\ “responsib*”) 
AND (“artificial intelligence” OR “AI” OR “machine learning”)\\ AND 
(“software developer” OR “software practitioner” OR “data scientist”\\
OR “machine learning” OR “software engineer” OR “programmer”)\\
\noalign{\smallskip}\hline\noalign{\smallskip}
\textbf{Final search string:} (``ethic*" OR ``moral*" OR ``fairness") AND (``artificial\\
intelligence" OR ``AI" OR ``machine learning" OR ``data
science") AND (``software \\developer" OR ``software practitioner" OR ``programmer")\\
\noalign{\smallskip}\hline
\end{tabular}
}
\end{table}

\subsection{Search} \label{search}
We performed the search using our \textit{seed review protocol}, presented in Table \ref{Table 3}. The search process was iterative and time-consuming because some combinations of search strings resulted in too many papers that were unmanageable to go through, whereas some combinations resulted in very few studies. Appendix \ref{Appendix B} contains the documentation of the search process showing the revision of the first search string through to the final search string.

\begin{table*}[htbp]
\centering
\caption{Seed and Final Grounded Theory Literature Review (GTLR) Protocols}
\label{Table 3}
\scriptsize
\begin{tabular}{>{\raggedright\arraybackslash}p{1.3cm}>{\raggedright\arraybackslash}p{4.6cm}>{\raggedright\arraybackslash}p{4.6cm}}
\hline\noalign{\smallskip}

 \multicolumn{1}{c}{\textbf{}}         & \multicolumn{1}{c}{\textbf{Seed GTLR Protocol}}                                         & \multicolumn{1}{c}{\textbf{Final GTLR Protocol}}\\
 \noalign{\smallskip}\hline\noalign{\smallskip}                               

 \multirow{4} {4em} {\textbf{Digital Databases}} & ACM Digital Library & No limitations on databases \\
& IEEE Xplore\\
& SpringerLink\\
& Wiley Online Library \\
\hline

\multirow {4} {4em} {\textbf{Search Items}} & Journal articles & Journal articles \\
& Conference papers & Conference papers \\
& Full Text & Students' theses\\
&  & Reports\\
& & Papers on arXiv\\
& & Full Text\\
\hline
\textbf{Language} & English & English\\
\hline
{\textbf{Publication Period}} & January 2010 to September 2022 & January 2010 to December 2022\\
\hline
\multirow{4} {2em} {\textbf{Search String}} & (``ethic*" OR ``moral*” OR ``fairness”) AND (``artificial intelligence" OR ``AI" OR ``machine & Snowballing applied in later iterations\\
&   learning" OR ``data science") AND \\
&  (``software developer" OR ``software \\ 
&  practitioner" OR ``programmer”)\\

\hline
\multirow {5} {4em} {\textbf{Inclusion Criteria}} & Each study must be a full-text published journal article or & Each study must be a full-text published journal article, \\
& conference paper & conference paper, students' thesis, report or paper on arXiv \\ & Studies that are written in English & Studies that are written in English \\
& Studies that are empirical & Studies that are empirical \\
& Studies that present AI practitioners’ views on ethics in AI & Studies that present AI practitioners’ views on ethics in AI \\

\hline
\multirow {6} {4em} {\textbf{Exclusion Criteria}} & Workshop articles, short papers (less than 4 pages), & Short papers (less than 4 pages) gray literature and incomplete \\ & books, gray literature, theses, unpublished and incomplete work & work\\
& Studies written in language other than English & Studies written in language other than English \\
& Review papers & Review papers \\
& Duplicate articles & Duplicate articles\\
& Theoretical or conceptual studies on ethics in AI (non-empirical) & Theoretical or conceptual studies on ethics in AI (non-empirical) \\
& AI related topics that do not include practitioners' perspectives & AI related topics that do not include practitioners' perspectives\\
\noalign{\smallskip}\hline
\end{tabular}%
\end{table*}

\subsection{Select}
We obtained a total of 1,337 primary articles (\textit{ACM DL: 312, IEEEX: 367, SpringerLink: 575 and Wiley OL: 83}) using the final search string (as shown in Table \ref{Table 2}) and the \textit{seed review protocol} (as shown in Table \ref{Table 3}). After filtering out the duplicates, we were left with 1073 articles. As per \citet{wolfswinkel2013using} grounded theory literature review (GTLR) guidelines, the next step was to refine the whole sample based on the title and abstract. We tried this approach for the first 200 articles each that came up in \textit{ACM DL, IEEEX, and SpringerLink} and all 83 articles in \textit{Wiley OL} to get a sense of the number of relevant articles to our research question. We read the abstracts of the articles whose titles seemed relevant to our research topic and tried to apply the inclusion and exclusion criteria to select the relevant articles. We quickly realised that selection based on title and abstract was not working well. This is because the presence of the key search terms (for example, ``\textit{ethics}'' AND ``\textit{AI}'' AND ``\textit{developer}'') was rather common and did not imply that the paper would include the practitioner's perspective on ethics in AI. We found ourselves having to scan through full texts to judge the relevance to our research question (RQ). Despite the effort involved, the return on investment was very low, for example, for every hundred papers read, we found only one or two relevant papers, i.e., those that included the AI practitioners' views on ethics in AI.

Out of 683 papers, we obtained only 13 primary articles that were relevant to our research topic. Many articles, albeit interesting, did not present the AI practitioners' views on ethics in AI. So, we decided to find more relevant articles through snowballing of articles. \textit{``Snowballing refers to using the reference list of a paper or the citations to the paper to identify additional papers"} \citep{wohlin2014guidelines}. Snowballing of those 13 articles via forward citations and backward citations was done to find more relevant articles and enrich the overview review quality. Snowballing seemed to work better for us than the traditional search approach. We modified the 
\textit{seed review protocol} accordingly, to include papers published in other databases and those published beyond journals and conferences, including students' theses, reports, and research papers uploaded to \textit{arXiv}. The \textit{final review protocol} used in this study is presented in Table \ref{Table 3}. In this way, we obtained 25 more relevant articles through snowballing, taking the total number of primary articles to 38. 

Here we note that the \textit{select} step of scanning through the full contents of 683 articles was very tedious with a very low return on investment, with only 13 relevant studies obtained. In hindsight, we would have done better to start with a set of \textit{seed} papers that were collectively known to the research team or those obtained from some quick searches on Google Scholar. What we did next by proceeding from the seed papers to cycles of snowballing, was more practical, productive, and in line with the iterative Grounded Theory (GT) approach as a form of applied theoretical sampling.

\subsection{Analyse}
Our review topic and domain lent themselves well to the socio-technical research context supported by socio-technical grounded theory (STGT) where our domain was AI, the actors were AI practitioners, the researcher team was collectively well versed in qualitative research and the AI domain, and the data was collected from relevant sources \citep{hoda2021socio}. We applied procedures of \textit{open coding}, \textit{constant comparison}, and \textit{memoing} in the basic stage and \textit{targeted data} collection and analysis, and \textit{theoretical structuring} in the advanced stage of theory development using the emergent mode.

The qualitative data included findings covered in the primary studies, including excerpts of raw underlying empirical data contained in the papers. Data were analysed iteratively in small batches. At first, we analysed the qualitative data of 13 articles that were obtained in the initial phase. We used the standard socio-technical grounded theory (STGT) data analysis techniques such as open coding, constant comparison, and memoing for those 13 articles, and advanced techniques such as targeted coding on the remaining 25 articles, followed by theoretical structuring. This approach of data analysis is rigorous and helped us to obtain multidimensional results that were original, relevant, and dense, as evidenced by the depth of the categories and underlying concepts (presented in Section \ref{findings}). The techniques of the socio-technical grounded theory (STGT) data analysis are explained in the following section. We also obtained layered understanding and reflections through reflective practices like memo writing, which are presented in Section \ref{discussion}. 

\subsubsection{The Basic Stage}
We performed open coding to generate \textit{codes} from the qualitative data of the initial set of 13 articles. Open coding was done for each line of the \textit{`Findings'} sections of the included articles to ensure we did not miss any information and insights related to our research question (RQ). The amount of qualitative data varied from article to article. For example: some articles had in-depth and long \textit{`Findings'} sections whereas some had short sections. Open coding for some articles consumed a lot of time and led to hundreds of codes whereas a limited number of codes were generated for some other articles.

Similar \textit{codes} were grouped into \textit{concepts} and similar 
\textit{concepts} into \textit{categories} using constant comparison. Examples of the application of Socio-Technical Grounded Theory (STGT)'s data analysis techniques to generate codes, concepts, and categories are shown in Figure \ref{Figure 3}, and a number of quotations from the original papers are included in Section \ref{findings}, to provide ``\textit{strength of evidence}'' \citep{hoda2021socio}. The process of developing concepts and categories was iterative. As we read more papers, we refined the emerging concepts and categories based on the new insights obtained. 
The coding process was initiated by the first author using Google Docs initially, and later, they transitioned to Google Spreadsheet due to the growing number of codes and concepts. Subsequently, the second author conducted a review of the codes and concepts generated by the first author independently. Following this review, feedback and revisions were discussed in detail during meetings involving all the authors. To clarify roles, the first author handled the coding, the second author offered feedback on the codes, concepts, and categories, while the remaining two authors contributed to refining the findings through critical questioning and feedback.

Each code was numbered as \textit{C1, C2, C3} and labeled with the paper ID (for example, G1, G2, G3) that it belonged to, to enable tracing and improve retrospective comprehension of the underlying contexts.

\begin{figure}[htbp]
    \centering
    \includegraphics [width=.6\linewidth]{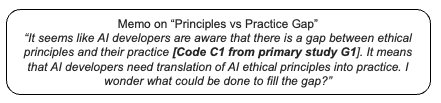}
    \caption{Example of a memo arising from the code (``principles vs practice gap") labeled [C1]} 
    \label{Figure 2}
\end{figure}

\begin{figure}[htbp]
    \centering
    \includegraphics [width=\linewidth]{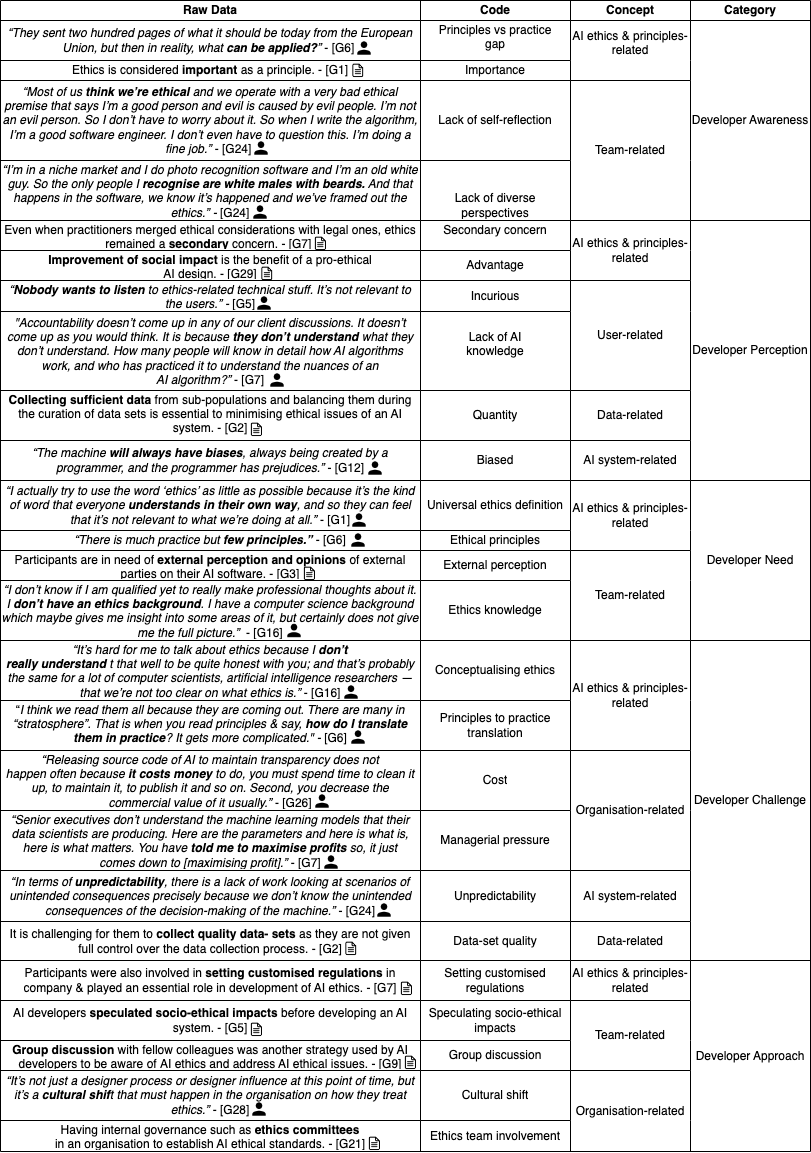}
    \caption{Example of Socio-Technical Grounded Theory (STGT) data analysis \citep{hoda2021socio} applied to primary studies ( \faUser : AI practitioner's quote; \faFileTextO : Literature data) } 
    \label{Figure 3}
\end{figure}

While the open coding led to valuable results in the form of codes, concepts, and categories, memoing helped us reflect on the insights related to the most prominent codes, concepts, and emerging categories. We also wrote reflective memos to document our reflections on the process of performing a grounded theory literature review (GTLR). These insights and reflections are presented in Section \ref{discussion}. An example of a memo created for this study is presented in Figure \ref{Figure 2}.

\subsubsection{The Advanced Stage} \label{targeted coding}
The codes and concepts generated from open coding in the basic stage led to the emergence of five categories: \textit{practitioner awareness}, \textit{practitioner perception}, \textit{practitioner need}, \textit{practitioner challenge} and \textit{practitioner approach} to AI ethics, with different level of details and depth underlying each. Once these categories were generated, we proceeded to identify new papers using forward and backward snowballing in the advanced stage of theory development. Since our topic under investigation was rather broad, to begin with, and some key categories of varying strengths had been identified, an \textit{emergent} mode of theory development seemed appropriate for the advanced stage \citep{hoda2021socio}. 

We proceeded to iteratively perform targeted data collection and analysis on more papers. Targeted coding involves generating codes that are relevant to the concepts and categories emerging from the basic stage \citep{hoda2021socio}. Reflections captured through memoing and snowballing served as an application of theoretical sampling when dealing with published literature, similar to how it is applied in primary socio-technical grounded theory (STGT) studies. 

We performed targeted coding in chunks of two to three sentences or short paragraphs that seemed relevant to our emergent findings, instead of the line-by-line coding, and continued with constant comparison. This process was a lot faster than open coding. The codes developed using targeted coding were placed under relevant concepts, and new concepts were aligned with existing categories in the same Google spreadsheet. In this stage, our memos became more advanced in the sense that they helped identify relationships between the concepts and develop a taxonomy. We continued with targeted data collection and analysis until all 38 selected articles were analysed. Finally, theoretical structuring was applied. This involved considering our findings against common theory templates to identify if any naturally fit. In doing so, we realised that the five categories together describe the main facets of how AI practitioners view ethics in AI, forming a form of multi-faceted taxonomy, similar to \citet{madampe2021faceted}.

\subsection{Present}
As the final step of the grounded theory literature review (GTLR) method, we present the findings of our review study, the five key categories that together form the multi-faceted taxonomy with underlying concepts and codes. We developed a taxonomy instead of a theory because we adhered to the principles outlined by \citet{wolfswinkel2013using} for conducting our Grounded Theory Literature Review and according to \citet{wolfswinkel2013using}, the key idea is to use the knowledge you've gained through analysis to decide how to best structure and present your findings in a way that makes sense and communicates your insights effectively. Likewise, we used the Socio-Technical Grounded Theory (STGT) method \citep{hoda2021socio} to analyse our data, which includes a recommendation: \emph{``STGT suggests that researchers should engage in theoretical structuring by identifying the type of theories that align best with their data, such as process, \textbf{taxonomy}, degree, or strategies \citep{glaser1978theoretical}.}" This is why we chose to create a taxonomy, as it was the most suitable approach based on the data we collected.

\begin{figure}[ht]
    \centering
    \includegraphics [width=\linewidth]{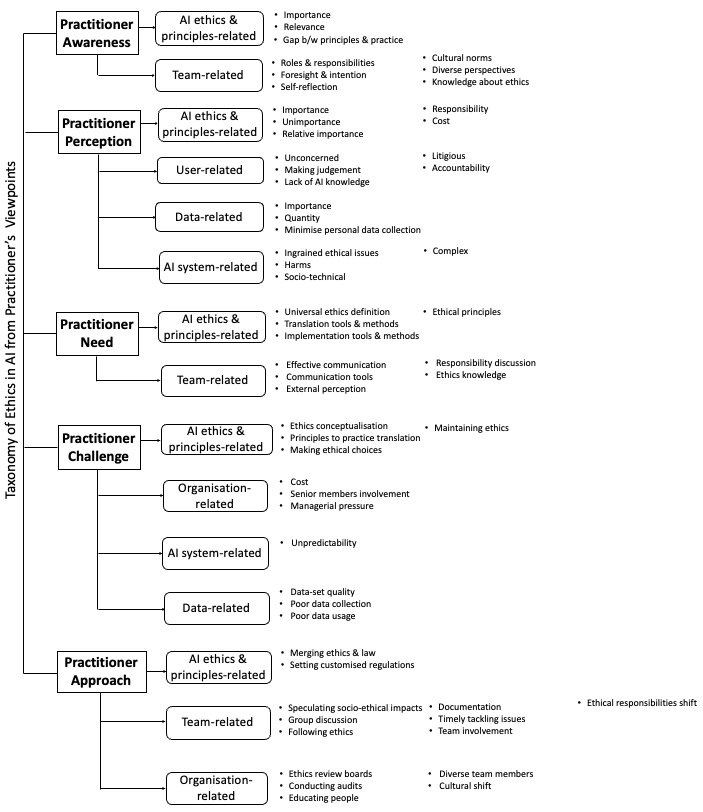} 
    \caption{Taxonomy of Ethics in AI from Practitioners’ Viewpoints}
    \label{fig:taxonomy}
\end{figure}

This is followed by a discussion of the findings and recommendations. In presenting the findings, we also make use of visualisations (see Figures \ref{fig:taxonomy} and \ref{fig:overview}) \citep{wolfswinkel2013using}. 

\section{Challenges, Threats and Limitations} \label{threats}
We now discuss some of the challenges, threats, and limitations of the Grounded Theory Literature Review (GTLR) method in our study. 

\textbf{Grounded Theory Literature Review (GTLR) nature}: Unlike a Systematic Literature Review (SLR), a Grounded Theory Literature Review (GTLR) study does not aim to achieve completeness. Rather, it focuses on capturing the `lay of the land' by identifying the key aspects of the topic and presenting rich explanations and nuanced insights. As such, while the process of a grounded theory literature review (GTLR) can be replicated, the results -- the resulting descriptive findings -- are not easily reproducible. Similarly, our study does not aim to be exhaustive, as it adheres to a grounded theory methodology. The chosen literature sample underwent thoughtful consideration, and although it is not all-encompassing, we have taken steps to assess its representativeness. Instead of using a \emph{representative sampling} approach, we used \emph{theoretical sampling} in our study, acknowledging that our sample might not exhibit the same level of representativeness as seen in a Systematic Literature Review (SLR), which is one of the limitations of our study.

\textbf{Search items and strategies}: Our search and selection steps for identifying the \textit{seed} papers and subsequent snowballing may have resulted in missing some relevant papers. This threat is dependent on the list of keywords selected for the study and the limitations of the search engines. To minimise the risk of this threat, we used an iterative approach to develop the search strings for the study. Initially, we chose the key terms from our research title and added their synonyms to develop the final search strings which returned the most relevant studies. For example, we included ``fairness" in our final search string because when we used only the term ``ethics", we obtained zero articles in two databases (\textit{ACM DL and Wiley OL}). The documentation of the search process is presented in Appendix \ref{Appendix B}. Likewise, we only used the term ``fairness" but did not include other terms like ``explainability" and ``interpretability" in our final search string. Due to this, there is a possibility that we missed papers that explore AI practitioners' views on these terms (``interpretability" and ``explainability"), which is a limitation of our study. 

The final search terms (“ethic*” OR “moral*” OR “fairness”) AND (“artificial intelligence” OR “AI” OR “machine learning” OR “data science”) AND (“software developer” OR “software practitioner” OR “programmer”) that we used in our study seem to be biased towards engineering/computer science publication outputs. This represents one of the limitations of our research since publications related to understanding AI practitioners' perspectives on `ethics in AI' may not exclusively reside within technical publications but may also extend to disciplines within the social sciences and humanities. Our use of these search terms, which are inclined towards outputs in engineering and computer science, might have led to the omission of relevant publications from social science and humanities domains.

In our final search query, we opted for the term ``software developer". Given the iterative nature of our keyword design process, we had previously experimented with incorporating keywords like ``data scientist", in combination with terms like ``AI practitioner" and ``machine learning engineer", to ensure that we did not inadvertently miss relevant papers. Unfortunately, this led to an overwhelming number of papers, posing a challenge for our study. Therefore, we decided to reduce the number of keywords and used only terms like ``software developer", ``software practitioner", and ``programmer" to obtain a more manageable set of papers for our study. However, we acknowledge that not including the term ``data scientist" in the search query may have caused us to miss some relevant papers, which is a limitation of our study.

The main objective of our study was to explore the empirical studies that focused on understanding AI practitioners’ views and experiences on ethics in AI. We were looking at the people involved in the technical development of AI systems but not managers, which is a limitation of our study. However, future studies could encompass managers, or separate reviews may delve into their perspectives on AI ethics. Likewise, we focused on studies published in the Software Engineering and Information Systems domains. However,  we acknowledge that AI practitioners' perspectives on AI ethics might have been extensively studied in social sciences and humanities, areas we didn't explore - a limitation of our study. Future research can encompass studies from these domains.

\textbf{Review protocol modification}: We decided to include only research-based articles in our grounded theory literature review (GTLR) study. Future grounded theory literature review (GTLR) studies can include literature from non-academic sources like in multi-vocal literature reviews (MLRs). Since, there is a lack of theories, frameworks, and theoretical models around this topic, we wanted to conduct a rigorous review study to present multidimensional findings and develop theoretical foundations for this critical and emerging topic. Finding enough empirical articles related to the research topic was another challenge. To overcome this, we had to make some adaptions to the original grounded theory literature review (GTLR) framework proposed by \cite{wolfswinkel2013using} and relaxed the review protocol during the snowballing of articles and included studies published in venues other than journals and conferences. We also used studies uploaded on arXiv as our \textit{seed} papers due to the lack of enough peer-reviewed publications relevant to our research topic. arXiv is a useful resource to find the latest research on emerging topics, and the quality of the work can be reasonably assessed from the draft. The growing impact of open sources like arXiv is evidenced by the increase in direct citations to arXiv in Scopus-indexed scholarly publications from 2000 to 2013 \citep{li2015role}.

\textbf{Time constraints}: We applied the socio-technical grounded theory \citep{hoda2021socio} approach to analyse the qualitative data of primary studies and focused on the \textit{`Findings'} section of the studies that presented empirical evidence. We did not find information on tools/software/framework/models used by AI practitioners to implement ethics in AI, although a study mentioned the existence of various tools but with no details provided [G10]. Since we were following a broad and inductive approach, we were not specifically looking for information on tools. This lack of information was surprising, but future reviews and studies can investigate the use of tools in implementing AI ethics.

\section{Findings} \label{findings}
As explained above, five key categories emerged from the analysis: (i) \textit{practitioner awareness}, (ii) \textit{practitioner perception}, (iii) \textit{practitioner need}, (iv) \textit{practitioner challenge} and (v) \textit{practitioner approach}. Taken together, they form \textit{a taxonomy of ethics in AI from practitioners' viewpoints}, shown in Figure \ref{fig:taxonomy}, with the underlying codes and concepts. Taken together, they represent the key aspects AI practitioners have been concerned with when considering ethics in AI. We describe each of the five key categories, and their underlying codes and concepts, and share quotes from the included primary studies by attributing them to paper IDs, \textit{G1} to \textit{G38}. The list of included studies is presented in Appendix \ref{Appendix A}.

\subsection{Practitioner Awareness} \label{awareness}
The first key category, or facet of the taxonomy, that emerged is \textit{Practitioner Awareness}. This category emerged from two underlying concepts: \textit{AI ethics \& principles-related awareness} and \textit{team-related awareness}.

\subsubsection{AI Ethics \& Principles---related Awareness} 
\label{AI ethics awareness}

The majority of articles reported that the developers participating in their study were aware of ethics, including its \emph{importance} [G1], [G4], [G5], [G17], and its 
\emph{relevance} [G18],[G8] in AI. Few studies [G1], [G23] reported that \emph{`Transparency'} was one of the ethical principles of AI that were discussed widely by the AI practitioners who participated in their studies and a highlighted topic of discussion in academia [G6]. Most AI practitioners who participated in studies [G9], [G18] were aware of the term `transparent AI' and it was recognised as a goal during AI development. \citet{mark2019ethics} [G17] mentioned that a participant was aware of the transparency law which helped them to determine what data needs to be public and what data needs to be private during the development of an AI system. Another participant in that study [G17] was also aware of transparency in AI and aimed at making transparent systems. The participant said: \faCommenting \hspace{0.05cm} \textit{``You might want to make it transparent for all citizens."}-- \textit{AI expert}--\textit{[G17]}.

Most AI practitioners who participated in a study [G2] were aware of the term \emph{`fairness'} which is an ethical principle of AI. Likewise, the majority of the participants of [G3] were aware of the importance of this principle and worked towards abolishing fairness-related issues in AI systems. Similarly, AI practitioners who participated in the studies [G3] and [G6] acknowledged that they were aware of the \emph{`accountability'} of AI systems and their importance, and nearly half of the participants (49\%) in a study [G4] felt responsible for the harm caused by their system. Similarly, some other studies supported the same idea. For instance, a study [G1] reported that one participant working on healthcare AI expressed a distinctive sense of responsibility compared to other respondents. They conveyed a more personalised accountability, feeling directly responsible for the well-being of certain users. Likewise, a study [G7] concluded that the participants of their study maintained responsibility as their specialised expertise not only enabled it but demanded it. A study [G8] reported \emph{`responsibility'} as an ethical principle that achieved the highest rank in terms of relevance in AI and it affected other ethical principles of AI. In this study, we use the terms `responsibility' and `accountability' interchangeably, following the definitions provided in Australia's AI ethics principles (See Appendix \ref{Appendix C} for clarification). In a study [G5], the participants reported that they were also aware that they possessed sensitive customer data so they actively considered accountability in relation to cyber-security and data management. \emph{`Privacy'} was another ethical principle that AI practitioners who participated in some studies were aware of and discussed widely. Privacy of data and information was identified as a major concern of organisations by some of the participants in a few studies including [G6], [G27]. A participant in [G6] said: \faCommenting \hspace{0.05cm}\textit{``And one of the first questions is privacy; that is, these algorithms that you are presenting, where are they going to be run? What will their information requirements be?"}-- \textit{AI practitioner}--\textit{[G6]}.
 
Few AI practitioners who participated in a study [G6] also seemed to be aware of the \emph{gap that exists between ethical principles and the practice} of implementing AI ethics. A participant in [G6] stated: \faCommenting \hspace{0.05cm}\textit{``They sent two hundred pages of what it should be today from the European Union, but then in reality, what can be applied? What is the reality of companies, and what is practical?"}-- \textit{AI practitioner}--\textit{[G6]}. Likewise, a few participants in other studies like [G15] and [G1] also agreed that there is a gap between academic discussion and industry practices.
 
\subsubsection{Team---related Awareness} \label{human awareness}
Participants in some studies acknowledged their awareness of their \emph{roles and responsibilities} in integrating ethics into AI during its development. For instance, a participant from the study [G7] highlighted being aware of their roles and responsibilities in implementing ethics during the development of AI systems. Similarly, a participant in another study [G23] expressed awareness of playing a pivotal role in shaping the ethics embedded in an AI system. 

Likewise, a participant in [G7] was also aware of his/her own limitations. The participant reported that sometimes the limitations of their \emph{foresight and intention} resulted in the development of a faulty and unethical AI system and stated: \faCommenting \hspace{0.05cm}\textit{``We are developing systems that are better than humans... only to discover as time goes on, that maybe they make things worse. And I don’t think that is a cynical thing to say. I think it is just a reflection of how every technological innovation has unfolded so far. What we need to do, as designers, is be aware that we could be designing the system that works and changes people’s lives, or you could be designing the system that makes people’s lives worse."} -- \textit{AI practitioner}--\textit{[G7]}. Likewise, a participant in another study [G8] and a participant in [G24] mentioned their inability to anticipate the unintended consequences of the machine's decision-making. The participant in [G24] stated: \faCommenting \hspace{0.05cm}\emph{``We don’t know the unintended consequences of the decision-making of the machine."}  -- \emph{AI engineer}-- [G24].

On the other hand, some studies highlighted the lack of such awareness and assumed ethical behavior, without addressing conscious and unconscious biases. For example: a study on practitioners' challenges in addressing ethical issues of AI presented various challenges that AI practitioners face in addressing AI ethical issues [G24]. The study [G24] reported that AI practitioners lacked \emph{self-reflection} in being able to recognise their own biases and responsibility which hampers AI ethics implementation. A participant in that study stated: \faCommenting \hspace{0.05cm}\textit{``Most of us think we’re ethical and we operate with a very bad ethical premise that says I’m a good person and evil is caused by evil people. I’m not an evil person. So I don’t have to worry about it. So when I write the algorithm, I’m a good software engineer. I don’t even have to question this. I’m doing a fine job."} -- \textit{AI engineer}--\textit{[G24]}. Similarly, several interviewees in a study [G2] highlighted the importance of taking into account biases ingrained in individuals at various phases of ML development, acknowledging the challenge of recognising their own biases.

Another participant mentioned that they lacked awareness about their \emph{cultural norms} and its impact while making ethical decisions during AI system development. \faCommenting \hspace{0.05cm}\textit{``The cultural norms that we have, but don’t even realise we have, that we use to make decisions about what’s right and wrong in context. It’s very difficult for any software system, even a really advanced one, to transcend its current context. It’s locked into however it was framed, in whatever social norms were in place amongst the developers at the time it was built."} -- \textit{AI engineer}--\textit{[G24]}. Similarly, a practitioner in [G25]  shared a similar idea about cultural norms and personal values when it comes to making ethical decisions.

Likewise, other participants stated that they did not always have a \emph{diverse and broad perspectives} to build inclusive AI technologies that affect the implementation of ethics in AI. \faCommenting \hspace{0.05cm}\textit{``I'm in a niche market and I do the photo recognition software and I’m an old white guy. So the only people I recognise are white males with beards. And that happens in the software, we know it’s happened and we’ve framed out the ethics."} -- \textit{AI engineer}--\textit{[G24]}. Similarly, participants in other studies like [G2], and [G21] were also aware of the importance of including diverse people in the team to ensure the ethical development of AI.

In a study [G25], a participant acknowledged their lack of \emph{knowledge about ethics} of AI. Similarly, participants in other studies, such as [G6] and [G8], also expressed awareness of their insufficient understanding of AI ethics and ethical principles.

\subsubsection{Overall Summary}
Few AI practitioners reported their awareness of the concept of AI ethics, ethical principles, their importance, and relevance in AI development. Likewise, very few AI practitioners were aware of the gap that exists between the ethical principles of AI and practice. Overall, this indicates a positive aspect concerning AI practitioners, as awareness of ethics is the initial step toward implementing ethical practices in AI development.

Similarly, some AI practitioners reported their understanding of the roles and responsibilities involved in the development of ethical AI systems. However, the primary focus of the majority of AI practitioners who participated in some studies was on recognising their own limitations that could result in the development of unethical AI systems. These limitations encompassed a lack of foresight and intention, insufficient self-reflection, limited knowledge of ethics, and a lack of awareness regarding cultural norms. In summary, this suggests that AI practitioners who participated in those studies engaged in significant introspection to comprehend the reasons behind the development of unethical AI systems. This introspective approach is positive because self-reflection can play a crucial role in identifying personal shortcomings and finding ways to address them.

\subsection{Practitioner Perception} \label{perception}
The second category is \textit{Practitioner Perception} which emerged from four underlying concepts: \textit{AI ethics \& principles-related perception}, \textit{User-related perception}, \textit{Data-related perception}, and \textit{AI system-related perception}. 

The perception category goes beyond acknowledging the existence of something and captures practitioners' views \& opinions about it, including held notions and beliefs. For example, it includes shared perceptions about the relative importance of ethical principles in developing AI systems, who is considered accountable for applying and upholding them, and the perceived cost of implementing ethics in AI. 

\subsubsection{AI Ethics \& Principles-related Perception} 
\label{AI ethics perception}
Perceptions about the importance of ethics varied. Some AI practitioners who participated in studies like [G1], [G29], and [G20] perceived `ethics' as very \textit{important} in developing AI systems. A study [G1] reported that AI practitioners acknowledged the importance of AI ethics. In the paper, when participants were asked if ethics is useful in AI, all (N=6) of them answered ``Yes".Nevertheless, it's important to consider that the participant sample size of this study [G1] was only 6.

In contrast, some AI practitioners participated in a few studies like [G7], [G38], [G9], and [G1] \emph{did not consider ethics as the important} element during AI system development. A study [G7] mentioned that AI practitioners who participated in their study considered only specific ethical principles important whereas another study [G9] mentioned that some practitioners in their study were less concerned about ethics as a whole in AI and more concerned about the usefulness and viability of their products. Participants in studies like [G7], and [G38] viewed `ethics' as a secondary concern, in [G1], it was seen as `other's problem,' and in the study [G10], a participant considered it a `non-functional requirement,' underscoring its \emph{unimportance} in AI development. A participant in a study [G38] stated: \faCommenting \hspace{0.05cm}\textit{``Ethics of AI and building AI responsibly is still not in the vernacular of your typical AI practice."} -- \textit{AI expert}--\textit{[G38]}. Likewise, a participant in [G34] shared similar thoughts on AI ethics. \faCommenting \hspace{0.05cm}\textit{``I don’t have time allocated during my normal week to think about responsible AI. This is not part of the work, at least not the part that someone would tell me from the top to worry about."} -- \textit{AI engineer}--\textit{[G34]}.

Developing responsible AI was seen as building positive relations between organisations and human beings by minimising inequality, improving well-being, and ensuring data protection and privacy. However, when it comes to the \emph{relative importance} of ethical principles, it was a divided house. An AI practitioner who participated in a study [G11] thought that AI systems must be fair in every way. Likewise, some participants in another study [G8] also thought that fairness issues in AI systems must not only be minimised but completely avoided, highlighting the importance of developing a fair AI system. On the other hand, within the same study [G8] surveying 51 participants, the highest importance, with an arithmetic mean of 4.71, was attributed to the principle of \emph{Protection of data privacy.} Other studies -- [G6] and [G10] -- also concluded that `Privacy protection and security' was the most important ethical principle in AI system development.

There were also differing opinions about who should be responsible (\emph{responsibility}) for ethics in AI. For example: in study [G30], a participant expressed uncertainty about the party responsible in the event of ethical incidents involving AI, mentioning that: \faCommenting \hspace{0.05cm}\textit{``When you think about who's accountable for AI that they're using in the public sector. When something goes bad, who do you point the finger at? If you got the human being out of the loop or maybe it's never out of the loop? But how do you decide who bears the cost of a bad experience?"} -- \textit{AI practitioner}--\textit{[G30]}. In contrast, others had strong opinions, such as a participant in [G10], who stated that ethics cannot be outsourced, suggesting it is ultimately the AI practitioners' responsibility. In a similar vein, there was a notion that AI practitioners are responsible for maintaining data privacy in AI systems. An AI practitioner in a study [G6] perceived the importance of privacy from the user's point of view and quoted: \faCommenting \hspace{0.05cm}\textit{``There you have the data of people, their addresses, you even have precious information, about when they are at home or not, private data, and making proper use of them is essential."}--\textit{AI practitioner}-- \textit{[G6]}. On the other hand, a participant in the same study [G6] perceived that both users and AI practitioners are responsible for maintaining the accountability of an AI system. Few participants in [G7] and [G26] also supported this idea. 

Another interesting opinion shared had to do with the perceived \emph{cost} of applying ethics in AI development. For example, too much ethical accountability was perceived as having a negative impact on business and organisational growth. A participant in [G6] stated: \faCommenting \hspace{0.05cm}\textit{``If I have to be very “ethical”, accuracy will also be affected. Then I think there is a dilemma there, at the end, of how ethical I am and how much business I am losing."}-- \textit{AI practitioner}--\textit{[G6]}. Likewise, the majority of participants in a study [G29] perceived the advantage of pro-ethical AI design as an improvement in social impact. However, a notable drawback mentioned was the associated costs, including resource costs and additional time.

\subsubsection{User-related Perception} \label{users perception}

Some AI practitioners who participated in studies like [G2], [G3], [G5], [G6], [G7], and [G34] had perceptions about users' nature, technical abilities, drivers, and their role in the context of ethics in AI. In this context, \emph{``users"} encompassed either the party commissioning a system, the end users, or both. We have provided additional clarity regarding the specific user categories that participants referred to when engaging in discussions about ethics in AI.

A participant in a study [G3] perceived that users only like to communicate if there is any chance of an incident occurring. Otherwise, they are \emph{unconcerned}. Both commissioning parties and end users were referred to as ``users" by the participants in [G3]. A similar perception was shared by a participant in another study [G5] who said that users are not curious about the workings of AI systems because the ethical technicalities of an AI system are irrelevant to them. The study participants didn't explicitly define the users in the study, but they referred to the individuals who would use the systems developed. These users could encompass both the clients commissioning the system and the end users who interact with the systems they create. Likewise, a participant in another study [G6] reported that users are concerned about ethics in AI and ethical issues only when it impacts their business, and a participant in [G34] mentioned that too much discussion of ethical AI could lead to users leaving who quoted: \faCommenting \hspace{0.05cm}\textit{``If you bring [ethical AI discussions] for every other use case and every other customer, there is already a lot of customers that we are losing. I don’t want this to create a bottleneck for our customers."} -- \textit{AI engineer}--\textit{[G34]}. The term \emph{``user"} in this study [G34] referred to the clients or companies who commission the AI product.

They also reported on the users' tendencies to \emph{judge} an AI model based on personal factors.  \faCommenting \hspace{0.05cm}\textit{``People tend to lose faith if their personally preferred risk indicators aren't in a model, even without looking at the performance of results."} -- \textit{ML practitioner}--\textit{[G3]}.

Clients' \emph{lack of AI knowledge} is one of the reasons that they have no interest in the ethics of AI according to a participant in a study [G7]: \faCommenting \hspace{0.05cm}\textit{``Accountability doesn't come up in any of our client discussions. It doesn't come up as you would think. It is because they don’t understand what they don’t understand. How many people will know in detail how AI algorithms work, and who has actually practiced it to understand the nuances of an AI algorithm?"} -- \textit{AI practitioner}--\textit{[G7]}. In this study, the term \emph{``clients"} denoted the people who commissioned their AI projects and they defined \emph{``clients"} as follows, \emph{``We refer to clients as those who commission and oversee AI projects, but do not do the technical work themselves"} [G7]. Likewise, a participant in [G5] also shared similar thoughts and said users don't want to listen to ethics-related stuff as they don't understand it. The participant in [G5] said, \faCommenting \hspace{0.05cm}\textit{``Nobody wants to listen to ethics-related technical stuff. It’s not relevant to the users."} -- \textit{AI developer}--\textit{[G5]}.

Participants in a study [G7] discussed the role of users in ethics in AI. An AI practitioner stated that it is essential to get users' needs and requirements before developing an AI system as it creates ethical parameters for them. Likewise, participants in a study [G7] perceived that the growth of an AI company is based on users. Users are likely to sue a company \emph{(litigious)} if the ethical issues of an AI system are not addressed by the company. \faCommenting \hspace{0.05cm}\textit{``[Companies] that aren't transparent or ethical, eventually, or you would hope, end up being prosecuted or sued or you know, all citizens as a whole would choose not to engage with them because they've been identified as an untrustworthy organisation. Because trust becomes the currency on which we trade. And will be more so as AI embeds itself in everything that we do."} -- \textit{AI practitioner}--\textit{[G7]}. The term \emph{``users''} in this study [G7] refers to the people who use AI systems. A similar thought was shared by some participants in [G2] who reported that they received customer complaints against the company if the customers faced any fairness-related issues with the products. The term `customer' in this study [G2] referred to end-users who used the products.

Similarly, some participants in a study [G7] perceived that users are equally \emph{accountable} as AI practitioners for the AI outcomes. A participant in that study stated: \faCommenting \hspace{0.05cm}\textit{``We were a technology provider, so we didn't make those decisions. It is the same as someone who builds guns for a living. You provide the gun to the guy who shoots it and kills someone in the army, but you just did your job and you made the tool."} -- \textit{AI practitioner}--\textit{[G7]}. This statement is supported by a participant in another study [G34] who said that users are equally \emph{accountable} for their own safety and quoted: \faCommenting \hspace{0.05cm}\textit{``I believe that the final responsibility lies at the client’s side who is finally deploying the actual service."} -- \textit{ AI engineer}--\textit{[G34]}. However, in a survey conducted by a study [G4], only 36\% of the respondents perceived that end users should take responsibility for their safety beyond what was explicitly outlined in the guidelines. 

\subsubsection{Data-related Perception}\label{data perception}
AI developers consider data as an \emph{important} aspect of implementing ethics in AI [G5], [G6]. A participant in a study [G5] perceived that data handling is an essential step that enhances the development of an ethical AI system. \faCommenting \hspace{0.05cm}\textit{``It's really important how you handle any kind of data that you preserve it correctly, among researchers, and don't hand it out to any government actors. I personally can't see any way to harm anyone with the data we have though."} -- \textit{AI developer}--\textit{[G5]}. 

The developer's na\"ive perception of the potential for harm (or lack thereof) is worth noting in the above example. Along with that, some participants in a study [G2] highlighted the importance of data collection and curation in AI system development. They mentioned that \emph{collecting sufficient data} from sub-populations and balancing them during the curation of data sets is essential to minimising the ethical issues of an AI system. A participant in [G15] also shared a similar idea on collecting sufficient ethical data for developing AI systems. 

On the other hand, some participants in a study [G18] reported that they \emph{minimised getting the personal data} of users or avoided its collection as much as possible so that no ethical issues related to data privacy arise during AI system development, whereas a participant in [G21] mentioned that they used privacy-preserving data collection techniques to reduce unethical work with data.  

\subsubsection{AI System-related Perception} \label{system perception} 
Some AI practitioners who participated in some studies like [G1], [G12], [G14], and [G34] perceived that AI systems have ingrained ethical issues. For example, a participant in a study [G1] perceived that every AI system \emph{has some ethical issues initially} and they take actions to either avoid or mitigate them. In a similar vein, a few participants in studies such as [G12], [G14], and [G34] reported that AI systems will always have some biases as humans create those systems. \faCommenting \hspace{0.05cm}\textit{``The machine will always have biases, always being created by a programmer, and the programmer has prejudices."} -- \textit{AI expert}--\textit{[G12]}. \faCommenting \hspace{0.05cm}\textit{``There is always a risk that the translation (AI) can be biased."} --\textit{AI engineer}--\textit{[G34]}.

Participants in various studies also compared and categorised the \emph{harms} of AI systems. For example, a participant in a study [G3] perceived the \emph{physical harms of an AI system as important and relevant} as compared to other harms and quoted: \faCommenting \hspace{0.05cm}\textit{``What could it affect the distribution of funds in a region, or could it result in a school taking useless action? It does have its own risks, but no one is going to die because of it."} -- \textit{ML practitioner}--\textit{[G3]}. Similarly, a participant in a study [G16] also perceived AI-based systems to be harmful and kill people: \faCommenting \hspace{0.05cm}\textit{``In my opinion, AI is going to kill people. Not in the way that everyone thinks it’s going to kill people, but people are going to die because of artificial intelligence. There is going to be job loss and it’s going to be rapid and rampant."} -- \textit{AI specialist} -- \textit{[G16]}

Some AI practitioners who participated in the studies [G7] and [G38] thought of AI as a \emph{socio-technical system} and not just a technical system. \faCommenting \hspace{0.05cm}\textit{``There is not really such a thing as an autonomous agent, it has kind of become important to say. It is now a socio-technical system, not just a technical system."} -- \textit{AI practitioner}--\textit{[G7]}. \faCommenting \hspace{0.05cm}\textit{``Responsible AI is a socio-technical concept. It’s not just like, using this library and implementing these algorithms, and suddenly your model is now fair and bias-free. It’s more so to think about the context of what your model is going to be deployed and where these harms originate and other things you can do."} -- \textit{AI expert}--\textit{[G38]}.

A participant in [G7] commented on the perceived limitations of AI systems, suggesting that they are so \emph{complex} that sometimes, they are not able to minimise ethical issues despite trying their best: \faCommenting \hspace{0.05cm}\textit{``I can say, yeah OK that was a fault, but this is how we did safety analysis. And I can see that this was missed, not because we were negligent, but just because it is so complicated. In this case, somebody died, but we did have the right ethical framework. But sometimes accidents happen. I think that is the kind of argument that you are going to have to make."} -- \textit{AI practitioner}--\textit{[G7]}. Similarly, a participant in [G2] shared a similar idea and emphasised that some ML systems are very complex and multi-component. Participants in another study [G9] also perceived AI systems as only complex concepts and prototypes so they did not feel accountable for the design of an AI system. 

\subsubsection{Overall Summary}
Overall, our synthesis says that AI practitioners who participated in the studies had both positive and negative perceptions about the concept of AI ethics. While some practitioners thought ethics were important to consider while developing AI systems, others perceived it as a secondary concern and non-functional requirement of AI. This diversity of views on AI ethics can have implications for the development and deployment of AI technologies and how ethical considerations are integrated into AI practices. Likewise, there were different views on the importance of different principles of AI ethics. Some practitioners perceived developing a fair AI is important whereas others perceived maintaining privacy during AI development is more important. This diversity in the views of different ethical principles might also impact the development of ethical AI-based systems. 

Perceptions regarding ethical considerations in the development of AI systems also extended to the question of responsibility. While some AI professionals felt it was their duty to create ethical AI systems and bear the accountability for any resulting harm, others believed that both users and practitioners shared this responsibility. We think it's essential to establish clear definitions of who should be accountable for ethical considerations during AI development and the consequences that arise from it. This way, there can be no evasion of this important issue. The discussion revolved around the expense associated with implementing ethical standards in AI development. We are curious whether, in the absence of cost barriers, AI practitioners could have created more ethically sound AI systems.

Some practitioners who participated in the studies also held unfavorable views regarding AI system users. Some believed that users generally did not pay much attention to AI ethics until actual ethical problems arose. Users were viewed as making judgments about AI systems based on personal biases rather than a deep understanding of how AI worked. Additionally, some participants perceived that users might resort to legal action against companies only when ethical issues with AI systems become apparent. Overall, this suggests a gap in user awareness and engagement with AI ethics, which could have implications for how AI is developed, used, and regulated.

Likewise, AI practitioners perceived a few steps to be important related to data to develop ethical AI systems. Proper data handling, sufficient data collection and data balancing, and avoiding personal data collection were perceived as important measures to mitigate ethical issues of AI systems. This implies that data-related practices contribute to ethical behavior and responsible AI development.

A few AI practitioners also had mixed perceptions about the nature of AI systems. Some expressed pessimism, suggesting that AI systems are excessively complex and inherently possess ethical issues that are difficult to mitigate. On the other hand, others viewed AI as socio-technical systems that, at the very least, take ethical considerations into account. Overall, this diversity in views highlights the ongoing debate and complexity surrounding AI ethics and underscores the importance of continued discussion and efforts to improve the ethical aspects of AI technology.

\subsection{Practitioner Need} \label{needs}
The review highlighted the different needs of AI practitioners which can help them enhance ethical implementation in AI systems. This category is underpinned by concepts such as \textit{AI ethics \& principles-related need} and \textit{team-related need}. 

\subsubsection{AI Ethics \& Principles---related Need} \label{ethics needs}

Practitioners in the included primary studies identified a number of needs. For example, the need for a \emph{universal ethics definition} was highlighted by the participant in a few studies, as it fulfills the gap between the ongoing academic discussion and the industry and enhances AI ethics implementation [G1], [G6], [G13]. A participant in [G1] said: \faCommenting \hspace{0.05cm}\textit{``I actually try to use the word ‘ethics’ as little as possible because it's the kind of word that everyone understands in their own way, and so they can feel that it's not relevant to what we're doing at all."} -- \textit{AI practitioner}--\textit{[G1]}.

Practitioners in [G1] and [G6] reported that participants expressed the need for \emph{tools or methods to translate principles into practice}. A participant in [G6] said: \faCommenting \hspace{0.05cm}\textit{``I think we read them all because they are coming out. There are many in the “stratosphere”. That is when you read the principles and say, How do I translate them in practice? It gets more complicated."}-- \textit{AI practitioner}--\textit{[G6]}. 

Likewise, a few AI practitioners who participated in a study [G1] and [G5] reported that they are challenged to implement ethics in AI as there is a lack of \emph{tools or methods for implementing ethics}. For example, in a study \textit{[G1]}, when AI practitioners were asked, \textit{``Do your AI development practices take into account ethics, and if yes, how?"}, all respondents (N=6) answered \textit{``No"}. This indicates that AI companies lack clear tools and methods that help AI practitioners implement ethics in AI. Another study [G19] concluded that there is a lack of tools that support continuous assurance of AI ethics. A participant in a study [G19] stated that it was challenging for them as they had to rely on manual practice to manage ethics principles during AI system development. 

All these points conclude that there is a need for tools that can help AI practitioners successfully implement ethics during AI system development. 
While the lack of practical tools is repeatedly identified, some participants in a study [G6] had an opposite view on the gap between principles and practice. They expressed the need for \emph{more principles} as they have much practice. \faCommenting \hspace{0.05cm}\textit{``There is much practice but few principles."} -- \textit{AI practitioner}--\textit{[G6]}.

\subsubsection{Team---related Need} \label{human needs}
There are a few needs related to AI practitioners that influence ethical implementation in a system. There is a need for \emph{effective communication} between AI practitioners as it supports ethics implementation [G2], [G3], [G15]. A few participants in studies [G2] and [G3] expressed the need for \emph{tools to facilitate communication} between AI model developers and data collectors. In the study [G2], out of those surveyed, 52\% of respondents (79\% of them when asked) expressed that tools aiding communication between model developers and data collectors would be incredibly valuable.

Similarly, some participants in a study [G3] reported that they are in need of \emph{external perceptions} and opinions of external parties on their AI software as well as AI ethical harms [G33]. It helps them to know the ethical issues of the software. A participant stated: \faCommenting \hspace{0.05cm}\textit{``For gender non-binary, we need to ensure we have the right people in the room who are experts on these harms and/or can provide authentic perspectives from lived experiences."} -- \textit{AI practitioner}--\textit{[G33]}.

On the other hand, a few participants in a study [G5], [G37] reported that they needed more \emph{discussion of their ethical responsibilities} in AI development as they were unsure about them. A participant in [G37] stated: \faCommenting \hspace{0.05cm}\textit{``It’s hard as when something is so new- we run into `Whose job is this?'"} -- \textit{AI practitioner}--\textit{[G37]}. However, Chivukula et al. (2020) [G28] reported that participants didn't feel responsible anymore as they were already doing their jobs ethically. \faCommenting \hspace{0.05cm}\textit{``I’m starting to feel like it’s not our responsibility anymore because I think all of us are already thinking from that perspective."} -- \textit{AI practitioner}--\textit{[G28]}. 

Similarly, some participants in [G18] reported that they were technology experts but didn't have any \emph{knowledge and background in ethics}. However, they were extremely aware of privacy concerns in AI use, highlighting an interesting relationship between practitioner awareness, perception, and challenges. A few participants in other studies like [G6], [G8], and [G25] also supported the notion. 

\subsubsection{Overall Summary}
The AI practitioners who participated in the included primary studies discussed several requirements concerning the conceptualisation of AI ethics and ethical guidelines. Some of them also expressed the necessity for tools and methodologies that could aid them in improving the development of ethical AI systems. This suggests that there is an ongoing need for support and resources to assist AI practitioners in adhering to ethical principles during the AI development process.

Similarly, a few participants in some of the included primary studies also addressed certain requirements regarding AI development teams. Some of these needs pertained to individual self-improvement, including the improvement of communication within the team and possessing a strong foundation in ethics as prerequisites for developing ethical AI systems. Additionally, there was a mention of the importance of discussing ethical responsibilities among team members as another requirement. Overall, the data suggests a commitment to improving the ethical aspects of AI development, both in terms of principles and practical implementation, and a recognition that addressing these ethical challenges requires a multifaceted approach involving teams and individual professionals.

\subsection{Practitioner Challenge} \label{challenge}
The fourth key category is \textit{Practitioner Challenge}. Several challenges are faced by AI practitioners in implementing AI ethics including \textit{AI ethics and principles-related challenge}, \textit{organisation-related challenge}, \textit{AI system-related challenge}, and \textit{data-related challenge}.

\subsubsection{AI Ethics \& Principles---related Challenge} \label{ethics challenge}
A number of challenges related to implementing AI ethics were reported, including knowledge gaps, gaps between principles \& practice, ethical trade-offs including business value considerations, and challenges to do with implementing specific ethical principles such as transparency, privacy, and accountability.

Participants in [G1] reported that they have difficulty in \emph{conceptualising ethics}, i.e., it is challenging for them to talk about ethics because the term `ethics' is understood differently by different people. A participant stated: \faCommenting \hspace{0.05cm}\textit{``I actually try to use the word ‘ethics’ as little as possible because it's the kind of word that everyone understands in their own way, and so they can feel that it's not relevant to what we're doing at all."} -- \textit{AI practitioner}--\textit{[G1]}. Some AI practitioners in [G34] expressed a similar notion, emphasising the need for more discussion on the practical application of AI ethics and the ethical consequences within the industry. They were worried that the absence of such discussions posed a challenge in grasping the concept of `ethics.' A participant stated: \faCommenting \hspace{0.05cm}\textit{``I think the whole issue of bias and its societal and ethical implications is terribly interesting and we don’t have as much conversation, particularly with cyber weapons, as we should."} -- \textit{AI engineer}--\textit{[G34]}.

Different types of challenges are mentioned and solutions are discussed in theory but there is no demonstration of those solutions in practice [G1], [G3]. \emph{Translation of AI principles into practice} is a challenge for AI practitioners as discussed by some participants in studies including [G1] and [G6].

AI practitioners are also challenged with \emph{making ethical choices} during the design of an AI system [G7]. \faCommenting \hspace{0.05cm}\textit{``Quite often we will make trade-offs naively and in line with our own experiences and expectations and fail to understand the implications of those trade-offs for others. We can assess all of the trade-offs, but we still don’t weigh them in impartial ways."} -- \textit{AI practitioner}--\textit{[G7]}. Some participants in studies like [G3] and [G22] supported the notion.

A number of challenges were mentioned to do with implementing specific ethical principles such as transparency, privacy, and accountability. For example, although transparency is perceived as an important ethical principle of AI, some AI practitioners who participated in studies like [G1], and [G6] faced challenges in \emph{maintaining transparency (ethics)}. These challenges arose both in the sense of transparency of systems and the development process [G1]. A participant in a  study [G6] mentioned that providing transparency to customers is challenging and quoted, \faCommenting \hspace{0.05cm}\textit{``There’s generally little transparency everywhere because it is hard to make that transparent to the customer I think it is still challenging to give that security and transparency."}-- \textit{AI practitioner}--\textit{[G6]}.  Similarly, a participant in a study [G22] was challenged to maintain accountability during AI development. \faCommenting \hspace{0.05cm} \textit{``How to clarify responsibilities and what are the standards or regulations? A machine cannot take responsibility by itself, as a human being can."} -- \textit{AI developer}--\textit{[G22]}.

\subsubsection{Organisation---related Challenge} \label{organisation challenge}
A study [G3] highlighted that communicating the performance of designed AI systems is challenging sometimes due to cost and business value considerations, which hampers the transparency of an AI system. \emph{Cost} is one of the major challenges in maintaining transparency in AI as reported by some participants in a study[G26]. \faCommenting \hspace{0.05cm}\textit{``Releasing source code of AI to maintain transparency does not happen often because it costs money to do, you have to spend time to clean it up, to maintain it, to publish it and so on. Second, you decrease the commercial value of it usually."} -- \textit{AI scientist}--\textit{[G26]}. 
Likewise, a participant in [G33] explained how a budget can be a challenge in developing ethical AI-based systems. \faCommenting \hspace{0.05cm}\textit{``If anybody wants us to do additional testing, which requires additional data gathering or labeling of existing data, right now we don’t have any budget set aside for that, so we need to proactively plan."} -- \textit{AI practitioner}--\textit{[G33]}.

\emph{Senior members of the company are involved} in setting the priority of AI practitioners' work and making decisions as discussed by a few participants in a couple of studies [G7], [G37]. Due to this, AI practitioners faced challenges such as \emph{communication issues and imbalance between AI practitioners and users}. A few participants of [G7] and [G37] reported: \faCommenting \hspace{0.05cm}\textit{``Senior executives don’t understand the machine learning models that their data scientists are producing. Here are the parameters and here is what is actually, here is what matters. You have told me to maximise profits so, it really just comes down to [maximising profit]."} -- \textit{AI practitioner}--\textit{[G7]}. \faCommenting \hspace{0.05cm}\textit{``More senior people are making the decisions. I saw ethical concerns but there was difficulty in communicating between my managers and my [responsible AI] team. People weren’t open for scrutinisation."} -- \textit{AI practitioner}--\textit{[G37]}.

Some AI practitioners who participated in a few studies faced \emph{managerial pressure} during AI development that influenced their ethics implementation [G33], [G35], [G37]. A participant in [G33] and another participant in [G35] quoted the following: \faCommenting \hspace{0.05cm}\textit{``We don’t have the luxury of saying, ‘Oh, we are supporting this particular locale \& this particular language in this particular circumstance.’ No, no, no, we’re doing it all! We’re doing it all at once, and we are being asked to ship faster. That is the pressure and there will be tension for anything that slows that trajectory because the gas pedal is to the metal."} -- \textit{AI practitioner}--\textit{[G33]}. \faCommenting \hspace{0.05cm}\textit{``There is always a time constraint in real work."} -- \textit{ML practitioner}--\textit{[G35]}.

\subsubsection{AI System---related Challenge} \label{system challenge}
The nature of AI-based systems creates challenges for AI practitioners while implementing ethics. The \emph{unpredictability} of an AI system was a major challenge for some AI practitioners who participated in a number of studies including [G1], [G4], and [G7] and they took actions to avoid, mitigate, or prevent unpredictable behaviors that took place [G1]. A participant in a study [G24] said: \faCommenting \hspace{0.05cm}\textit{``In terms of unpredictability, there is a lack of work looking at scenarios of unintended consequences precisely because we don’t know the unintended consequences of the decision-making of the machine."} -- \textit{AI engineer}--\textit{[G24]}. Some external causes of AI system unpredictability were also discussed by a few participants in a study [G1] such as cyber-security threats. Likewise, clients' needs such as profit maximisation and attention optimisation were mentioned as one of the causes of unpredictable system behavior that ultimately develops ethical issues. A participant in a study [G7] stated: \faCommenting \hspace{0.05cm}\textit{``It’s not that we thought what we were doing was safe, it’s just that, certain inbuilt desires to increase clicks, to increase attention, to maximise advertising was our primary motivation. You did not have to think about any other consequences."} -- \textit{AI practitioner}--\textit{[G7]}. Certain participants in a study [G4] also deliberated on the challenges associated with addressing the unpredictable behaviour of AI systems.  Not all organisations and their AI practitioners have fallback plans for solving ethical issues developed by an unpredictable AI system [G4]. According to a survey conducted by a study [G4], nearly half (48\%) of AI practitioners mentioned that their companies lacked contingency plans if the AI systems they develop show unpredictable behaviors. Therefore, it can be concluded that it is challenging for AI practitioners to solve ethical issues that are developed by unpredictable AI system behaviors. 

\subsubsection{Data---related Challenge} \label{data challenges}
Some of the challenges shared by participants across the primary studies were related to data. For example: the \emph{quality of the data set} used in AI algorithms was considered one of the main factors affecting the fairness of an AI system by some participants in a study [G37]. Likewise, a few AI practitioners in [G2] mentioned that it was challenging for them to collect quality data sets as they were not given full control over the data collection process. It was supported by other studies like [G17] and  [G33] as one of the participants in [G33] stated: \faCommenting \hspace{0.05cm}\textit{``We barely have access to data-sets, to begin with, so we take anything that we can get basically"} -- \textit{AI practitioner}--\textit{[G33]}.

Similarly, AI practitioners involved in some primary studies including [G2], [G22], and [G33] found that challenges arose from \emph{poor data collection} processes in AI development, stemming from insufficient user engagement with the product [G2]. Some participants in a study [G2] also mentioned that challenges to getting additional training data to ensure AI fairness arose due to the team's blind spots. According to Holstein et al. (2019) [G2], participants reported cases in which AI systems recognised celebrities in some countries but not others. \faCommenting \hspace{0.05cm}\textit{``It sounds easy to just say like, `Oh, just add some more images in there,’ but there’s no person on the team that actually knows what all of [these celebrities] look like if I noticed that there’s some celebrity from Taiwan that does not have enough images in there, I actually don’t know what they look like to go and fix that. But Beyonc\'{e}, I know what she looks like."} -- \textit{ML practitioner}--\textit{[G2]}.
Likewise, a few participants in a study [G33] were unable to evaluate the fairness of the AI-based system they developed due to the \emph{lack of proper data collection} methods which was one of the challenges they faced. A participant stated: \faCommenting \hspace{0.05cm}\textit{``So I guess I’m just having trouble getting over the hurdle that I don’t think we have a real approved data collection method [for data that lets us evaluate fairness] at all."} -- \textit{AI practitioner}--\textit{[G33]}.

On the other hand, in some cases, data privacy issues were seen to induce risk aversion and impose barriers to better \emph{data usage}. A participant in a study [G6] quoted: \faCommenting \hspace{0.05cm}\textit{``My perception is that companies do take great care of their information, to the point that they often prefer not to generate value from information [rather] than to expose their information to a risk of leakage."} -- \textit{AI practitioner}--\textit{[G6]}. Similar thoughts on the use of data were shared by a participant in another study [G12].  

\subsubsection{Overall Summary}
Participants in the included primary studies discussed various challenges related to the concept of AI ethics and ethical principles. Some participants discussed challenges related to ethics, including variations in how people understand ethics, the practical application of ethical principles, and the consistent adherence to various ethical standards throughout the AI development process. In general, this data suggests that the primary challenge for practitioners is grasping the essence of ethics, which we consider to be the fundamental issue and should be prioritised for resolution. 

Similarly, organisations have contributed to obstructing AI practitioners in their efforts to develop ethical AI systems. Challenges raised by participants, such as limited budgets for integrating ethics, tight project deadlines, and restricted decision-making authority during AI development, indicate that organisations could assist AI practitioners by addressing these issues when feasible.

Some participants also discussed the challenges regarding the unpredictability of AI systems. They identified factors contributing to this unpredictability, such as profit maximisation, attention optimisation, and cyber-security threats. The absence of contingency plans to address issues stemming from AI system unpredictability was also discussed. Overall, it indicates that AI practitioners employ certain strategies to mitigate unpredictability in AI systems, but there is a demand for methods and tools to effectively prevent or manage such unpredictability. The development of such methods or tools would aid in reducing ethical risks associated with AI.

Participants discussed challenges associated with the data used to train AI models. They explained how the quality of data and the processes involved in handling data can influence AI development. Some AI practitioners faced challenges related to ensuring the ethical development of AI, primarily due to issues like inadequate data quality, poor data collection practices, and improper data usage. Overall, the data suggests that to ensure ethical AI development, it is essential to address issues related to data quality and data handling processes.

\subsection{Practitioner Approach} \label{approach}
The review of empirical studies provided insights into the approaches used by AI practitioners to implement ethics during AI system development. This category is underpinned by three key concepts, \textit{AI ethics \& principles-related approach}, \textit{team-related approach}, and \textit{organisation-related approach} to enhance ethics implementation in AI. AI practitioners discussed the applied and/or potential strategies related to these three concepts. Applied strategies refer to the techniques or ways that AI practitioners reported using to enhance the implementation of ethics in AI, whereas possible strategies are the recommendations or potential solutions discussed by AI practitioners to enhance the implementation of ethics in AI.

\subsubsection{AI Ethics \& Principles---related Approach} \label{ethics-strategies}
AI practitioners discussed the applied strategies related to AI ethics and ethical principles. For example, they reported \emph{merging ethical and legal considerations} to ensure no illegal actions have been taken during AI system development [G7]. In this strategy, ethics remained a secondary concern. A participant stated: \faCommenting \hspace{0.05cm}\textit{``The very minimum that you have to adhere to is the law. So, we start by ensuring that everything that we do, or our clients do is legal. Then we have to decide whether or not it is appropriate, which could be considered ethical or fair."} -- \textit{AI practitioner}--\textit{[G7]}. Similarly, in a study [G4], some participants noted that existing laws, like the General Data Protection Regulation (GDPR), compel them to address ethical concerns related to AI. This suggests the potential synergy of integrating laws and ethics to promote the ethical advancement of AI systems. 

AI practitioners were also involved in \emph{setting customised regulations} in the company and played an essential role in the development of AI ethics. This strategy was used to enhance ethics implementation by developing comprehensive and well-defined guidelines for AI ethics for the company [G7]. Some participants in a study [G11] also reported that they needed to customize the general policies in the organisation to better support privacy and accessibility for their specific circumstances to ensure AI fairness. 

\subsubsection{Team---related Approach} \label{team-strategies}
Some participants in a study [G1] reported that organisations used proactive strategies such as \emph{speculating socio-ethical impacts} and analysing hypothetical situations to enhance ethics implementation in AI development. Likewise, a few participants in another study [G5] supported the notion and mentioned that such strategies aimed to address ethical issues that may arise and plan for their potential consequences [G5]. Analysing a hypothetical situation of unpredictability was a strategy used to solve an AI system's unpredictable behavior [G1]. Similarly, a participant in a study [G2] reported that speculating possible fairness issues of an AI system before deploying it was a strategy used to minimise fairness issues [G2] in AI.

AI practitioners also used \emph{group discussions} with colleagues [G9] and sought information from secondary sources like blog posts and videos and primary sources like academic papers [G32] to stay informed about AI ethics and address ethical issues. Similarly, a participant in a study [G17] mentioned that they had an interaction and collaborative discussion with policymakers and legal teams of the company to ensure that their algorithms were abiding by the legislation. This denotes that AI companies focused on ensuring their algorithms were legally fit before deployment. Likewise, AI practitioners also consulted with domain experts and relevant stakeholders during the data analysis phase of AI development. A participant stated: \faCommenting \hspace{0.05cm}\textit{``We consult data-set builders about how the data was collected and how the features [were] being defined."} -- \textit{ML practitioner}--\textit{[G35]}.

Some participants in studies including [G18] and [G31] reported that they \emph{followed codes of ethics and standards of practice} while developing AI-based systems and stated: \faCommenting \hspace{0.05cm}\textit{``We follow regulations, but since our software is not a very risky one, we haven’t taken much caution."} -- \textit{AI specialist}--\textit{[G31]}.

A few AI practitioners who participated in the included primary studies discussed some proactive strategies and methods that they used to maintain transparency and accountability of AI systems. \emph{Documentation} of the codes was the primary proactive strategy for creating transparency during the development of an AI system and tracking the actions and people involved as discussed by some participants in studies including [G15], [G31]. Similarly, documenting decisions made by AI practitioners to track decisions back to individuals when needed was one of the strategies used to enhance accountability by a few participants in studies like [G10], and [G31]. A participant in [G31] said: \faCommenting \hspace{0.05cm}\textit{``These factors have now been added to our developmental process because it has been seen that it is an advantage to increase our customers and our knowledge about these, especially when facing new challenges with the near future AI technology. It is good to prepare in advance, not after something has happened."-- \textit{AI practitioner}--\textit{ [G31]}}.

However, some companies did not use proactive strategies to maintain transparency of AI systems but addressed transparency issues only when it impacted their business [G6]. Some AI practitioners just followed what is legal and \emph{shifted the ethical responsibilities} to policymakers and legislative authorities [G7]. In contrast, some participants in a study [G24] placed the ethical responsibility on the company manager.

In addition to sharing experiences of tried and tested strategies, practitioners also discussed potential strategies that they thought could improve ethics in AI. A study [G10] concluded that appointing one individual to implement ethics during AI development is not a good option. The whole \emph{AI development team must be involved} in the process of ethics implementation. In another study [G15], a participant proposed a similar notion, emphasising the involvement of not just senior members but also junior AI practitioners in integrating ethics during AI development.

Likewise, a participant in a study [G10] mentioned that \emph{tackling ethical issues timely} i.e., during the design and development of an AI system to enhance system transparency is good. In another study [G4], a participant recommended addressing ethical concerns during the development of AI systems, highlighting the necessity for providing AI developers with supportive methods. 

\subsubsection{Organisation---related Approach}
Some participants in a study [G18] reported several strategies provided by organisations to enhance ethics implementation in AI such as \emph{ethics review boards}. Likewise, a participant in a study [G21] mentioned that having internal governance such as ethics committees in an organisation to establish AI ethical standards can provide AI practitioners an opportunity to work closely with ethicists so that they can verify if ethics is being implemented appropriately during AI system development.

Some participants in studies like [G1] and [G5] stated that \emph{conducting audits} was the other important strategy organisations provided to them to solve transparency issues. A participant in [G21] reported that employing AI auditors could help AI practitioners in developing ethical AI systems. 

\emph{Educating people} i.e., practitioners about AI ethics to help them become aware of AI ethics and ethical issues was a potential strategy discussed by some participants in studies like [G21], [G37], and [G38]. A participant in [G38] stated: \faCommenting \hspace{0.05cm}\textit{``Providing an e-learning program to all employees is important and such programming includes what AI ethics is, and why that matters and what kind of incidents actually happen in the market"} -- \textit{AI expert}--\textit{[G38]}. However, a participant in a study [G28] stated that educating business owners with ethics training and education instead of them because they focus on their business growth rather than ethics in AI. The participant said: \faCommenting \hspace{0.05cm}\textit{``More education for business owners and people in other parts of businesses to be responsible business owners. Don’t push these agendas. You think making more money quickly is the most important part of your business."} -- \textit{AI practitioner}--\textit{[G28]}.

Some AI practitioners who participated in some of the included primary studies also discussed some potential strategies that organisations should provide them to help them enhance ethics implementation during AI development. \emph{Including diverse team members} in the development team was one of them [G21], [G36]. A participant in [G36] stated: \faCommenting \hspace{0.05cm}\textit{``No one in the developing team speaks the language and knows the idioms — how would they properly audit the outcomes? That’s why it is a good idea to spend time bringing native speakers into the auditing process."} -- \textit{AI practitioner}--\textit{[G36]}. Similarly, hiring employees who belong to different communities and ethnic groups was reported as a potential strategy to enhance the chance of spotting biases within a team by some participants in the included primary studies. A participant in a study [G2] suggested using fairness-focused quizzes in the interview processes can be useful for hiring people who can detect fairness issues in an AI system [G2]. \faCommenting \hspace{0.05cm}\textit{``No one person on the team [has expertise] in all types of bias especially when you take into account different cultures. It would be helpful to somehow pool knowledge of potential fairness issues in specific application domains across teams with different backgrounds, who have complementary knowledge and blind spots."} -- \textit{ML practitioner}--\textit{[G2]}.

Similarly, a participant in a study [G28] mentioned that organisations should work on treating ethics properly by \emph{having a cultural shift} in the organisation: \faCommenting \hspace{0.05cm}\textit{``It’s not just a designer process or designer influence at this point of time, but it’s a cultural shift that has to happen in the organisation on how they treat ethics."} -- \textit{AI practitioner} --\textit{[G28]}. Having a cultural change in the company was discussed by a participant in another study as well [G4].

\subsubsection{Overall Summary}

Participants discussed several strategies that they used to ensure the ethical development of AI systems. The applied strategies related to AI ethics and principles were used by the participants to ensure the ethical development of AI systems such as merging ethics and law and setting customised AI ethics regulations in the company. Overall, this indicates that practitioners emphasize the comprehensive integration of all AI ethical principles to ensure that no aspect is overlooked during the development process. 

Some approaches were performed by the team to ensure the ethical development of AI systems such as group discussions with colleagues on AI ethics, analysing hypothetical situations of AI ethical issues, considering socio-ethical impacts of AI, and discussion with policymakers and legal teams to ensure algorithms are abiding by laws. Overall, this data suggests a comprehensive and multidisciplinary approach to addressing AI ethics, where the team actively engages in discussions, analysis, and collaboration with various stakeholders to promote the ethical development of AI systems.

Some participants mentioned that their organisations currently use various methods, such as audits, and ethics review boards, to promote ethical AI development. However, the discussion highlighted a greater emphasis on potential approaches that organisations could offer to their AI development teams to ensure ethical AI. For instance, some participants proposed that organisations could prioritise diversity within AI teams, provide education and training on AI ethics for practitioners, establish internal governance mechanisms like ethics committees, cultivate a cultural shift within the organisation towards ethical considerations, and implement tools like quizzes during the hiring process for AI teams to enhance ethical development. It indicates that organisations can offer additional support to AI practitioners in their pursuit of ethical AI systems, suggesting that there is more that can be done in this regard.

\section{Discussion and Recommendations} \label{discussion}

\subsection{Taxonomy of Ethics in AI from Practitioners' Viewpoints.} \label{taxonomy}
\begin{figure}[htbp]
    \centering
    \includegraphics [scale=0.35]{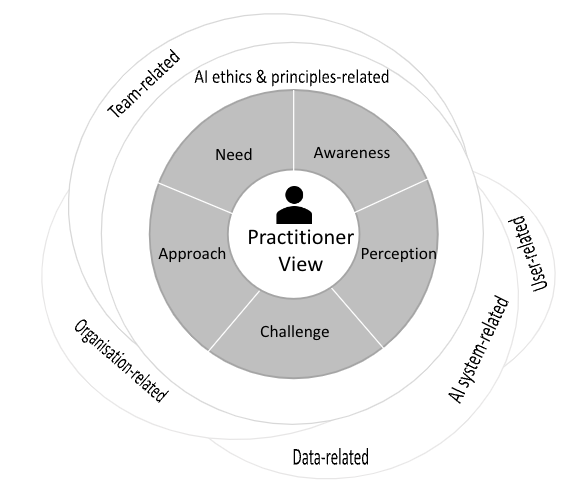} 
    \caption{An Overview of the Aspects of Ethics in AI from AI Practitioners' Viewpoints} 
    \label{fig:overview}
\end{figure}

The \textit{taxonomy of ethics in AI from practitioners' viewpoints} aims to assist AI practitioners in identifying different aspects related to ethics in AI such as their awareness of ethics in AI, their perception towards it, the challenges they face during ethics implementation in AI, their needs, and the approaches they use to enhance better implementation of ethics in AI. Using the findings, we believe that AI development teams will have a better understanding of AI ethics, and AI managers will be able to better manage their teams by understanding the needs and challenges of their team members.

An overview of the taxonomy and the coverage of the underlying concepts across the categories is presented in Figure \ref{fig:overview}. As mentioned previously, we obtained multiple concepts for each category. Some concepts were common across some categories whereas some were unique. For example, \textit{`AI ethics \& principles'} is a concept that emerged for each of the five categories, depicted by a full circle around the five categories. The `\textit{teams-related}' concept emerged for three categories, namely, \textit{practitioner awareness}, \textit{practitioner need}, and \textit{practitioner approach}, depicted by a crescent that covers these three categories on the top left. While the `\textit{user-related}' concept emerged for only one category, \textit{practitioner perception}, as seen by a small crescent over that category. The codes underlying these concepts were unique to each category, as seen in Figure \ref{fig:taxonomy} and described in the \textit{`Findings'} section.

The overview of the taxonomy shows that AI practitioners are mostly concerned about AI ethics and ethical principles. For example, they discussed their \textit{awareness} of ethics [G16] and different AI ethical principles such as transparency [G17], accountability [G3], fairness [G2], and privacy [G6] and also shared their positive \textit{perception} such as its importance and benefits, and negative perceptions such as the high cost of ethics application [G6] and ethics being a non-functional requirement in AI development [G10]. Likewise, they mentioned different \textit{challenges} they faced during AI ethics implementation which are related to AI ethics and principles such as ethics conceptualisation [G1], the difficulty of translating principles to practice [G6] and making ethical choices [G7]. Their \textit{needs} related to AI ethics and principles were also reported by AI practitioners in the literature including the need for universal ethics definition [G1], tools to translate principles to practice [G6] along with the \textit{approaches} they used related to AI ethics and principles to enhance better implementation of ethics in AI such as merging ethical and legal considerations and setting customised regulations in the organisation [G7].

On the other hand, the review shows that AI practitioners have been less concerned about the aspects related to \textit{users} when it comes to ethics in AI. For example: AI practitioners perceive that users are unconcerned and incurious [G5] about the ethical aspects of AI software they use unless there is any chance of an incident occurring [G3]. Likewise, they reported that users don't have much knowledge about AI which makes them uninterested in the ethical aspects of AI-based systems [G7]. No challenges or needs related to users were reported in the literature that impact AI practitioners' AI ethics implementation in AI-based systems. In conclusion, \textit{AI ethics and principles} and \textit{team-related aspects} were front and center for AI practitioners while they lacked a better view of the \textit{user-related} aspects. Our findings contribute to the academic and practical discussions by exploring the studies that have included the views and experiences of AI practitioners about ethics in AI. As we conducted a grounded theory literature review (GTLR), we got an opportunity to rigorously review the primary empirical studies relevant to our research question and develop a taxonomy. We now discuss some of the insights captured through memoing and team discussions, accompanied by recommendations.

\subsection{Ethics in AI -- Whose Problem is it Anyway?} 
Participants of the primary studies had different perceptions of AI ethics and its implementation. Most studies included in our research concluded that AI practitioners perceived ethics as an essential aspect of AI [G5], [G20]. However, some participants had other viewpoints. A participant in [G1] stated that discussion on AI ethics does not affect most people, except for AI ethics discussions in massive companies like Google. Another participant from [G4] perceived ethics as a non-functional requirement in AI, something to be implemented externally [G23]. In contrast, a participant in [G4] stated that ethics could not be ``outsourced", and it should be implemented by AI practitioners who are developing the software. The diverse perspective of the participants about the implementation of ethics in AI serves to highlight the complex nature of the topic and why organisations struggle to implement AI ethics.

Likewise, there were also different views on who should be accountable for implementing ethics in AI. An AI practitioner in a study [G30] shared the uncertainty typically present when deciding who or what is responsible when ethical issues arise in AI systems. It seems certain organisations attempt to define who should be held accountable, but again, there is no universal understanding. For example: the \textit{ACM Code of Ethics} clearly puts the responsibility on professionals who develop these systems. On the other hand, AI practitioners perceive that only physical harm caused by AI systems is essential and needs to be considered [G3]. This statement is alarming as it hints that some practitioners carry the view that only physical harm is worth being concerned about.

\noindent \textit{Recommendations for Practice}
\begin{itemize}
    \item[\faThumbsUp] 
    Given the diverse perspectives on who owns accountability for considering ethics in AI systems development and potential ethical issues arising from AI system use, it is important for AI development teams, which are usually multidisciplinary in nature, as well as managers and organisations at large to have open discussions about such issues at their workplace [G5]. The lack of discussion about ethics within the tech industry has been identified as a significant challenge by engineers  \citep{metcalf2019owning}. For example, this can be done through organising discussion panels, guest seminars by ethics and ethical AI experts, and hosting open online forums for employees to discuss such topics. Another approach is to collate the challenges specific to the organisation and see how they map to selected ethical frameworks, as was conducted at Australia’s national scientific research agency (CSIRO) [G26]. 
    
    \item[\faThumbsUp] Practitioner discussions can be followed by strategic and organised attempts to reconcile perspectives, for example, teams collaboratively selecting an existing or creating a bespoke ethical framework, and drafting practical approaches to implement them in their specific project contexts [G7], many of which may be application domain specific.

    \item[\faThumbsUp] We recommend \textit{proactive awareness} as evidenced in our reviews, such as driven by personal interest and experiences [G6], organisational needs [G3], and regulations such as the General Data Protection Regulation (GDPR) [G6]. Whereas \textit{reactive awareness}, driven by customer complaints about AI ethical issues and negative media coverage [G2], is not desirable.
    
    \item[\faThumbsUp] Similarly, we recommend \textit{proactive strategies} such as \textit{speculating socio-ethical impacts} by AI practitioners prior to developing an AI system [G5]. \emph{Speculating socio-ethical impacts} hints at speculative design approaches which have been heavily discussed and supported by multiple studies as well \citep{lee2023speculating,alfrink2023contestable}.  \textit{Analysing hypothetical situation of unpredictability} to solve unpredictable behaviour of an AI system [G1], \textit{following codes of ethics and standards of practice} [G18], \textit{including diverse people} in the development team [G21], and having internal governance such as \textit{ethics committees} in an organisation to establish AI ethical standards [G21] are also other proactive strategies we recommend. 
    
    \item[\faThumbsUp] Finally, there is also a need to consider accountability at the organisation and industry levels. For example: Ibanez et al. (2021) [G6] reported that there is a need for ethical governance that can help them solve accountability issues.
\end{itemize}

\subsection{Ethics-critical Domains Lead the Way.}
Comparisons were made between the medical field and the IT field in terms of the awareness of ethical regulations in AI [G5]. Participants mentioned that practitioners developing AI used in the medical field are more aware of ethics because the medical field has stricter laws and regulations than IT. This hints that awareness of AI ethics depends on \textit{domain specificity}. Domains such as medical and health are more ethics-aware than others and lead the way in ethics awareness and implementation.

\noindent \textit{Recommendations for Practice}
\begin{itemize}
    \item[\faThumbsUp] The IT domain can learn from the advances in improving the awareness of and implementing ethics in the medical domain \citep{mittelstadt2019principles}. This includes digital, virtual, mobile, and tele-health areas, as well as AI systems developed in other domains.
    \item[\faThumbsUp] Labelling certain domains as safety-critical and equating that with ethics-critical, can be a flawed argument leading to perceptions that domains traditionally considered non-safety-critical, such as gaming and social media, can be held to lower standards and expectations when it comes to ethics implementation. We know from multiple cases of cyberbullying and `intelligent' games encouraging self-harm in young adults (for example, \emph{`The Blue Whale Game'} \citep{mukhra2019blue}) that this would be a mistake. We recommend that all domains should aim to be ethics-critical. 
    
\end{itemize}

\subsection{Research Can Help in Fundamental and Practical Ways.}
The perspectives of AI practitioners on the nature of AI systems can have a significant impact on the implementation of ethics in AI. Some practitioners may view AI as a socio-technical system and therefore place a strong emphasis on ethics [G4], while others may view AI as a complex system and find it challenging to address ethical issues, leading them to avoid ethical considerations [G7]. The participants' perspectives on AI systems indicate that the implementation of ethics depends on how practitioners perceive AI ethics.\\

\noindent \textit{Recommendations for Research}

\noindent Based on our review findings, we recommend research including empirical studies, reviews, and solutions \& tools development into the following topics.
\begin{itemize}
    \item[\faThumbsUp] Most of the participants in a study [G9] reported that there is no use of ethical tools in AI companies to enhance ethics implementation in AI. Therefore, reviewing tools available to AI practitioners to enhance the AI ethics implementation including their evaluation and feedback for improvement would be helpful to make them aware of the tools that are beneficial.

    \item[\faThumbsUp] Based on our findings, it appears that some AI practitioners involved in studies such as [G5, G6, G19] mentioned the need for assistance in the form of tools and methodologies to effectively integrate ethics into AI and put ethical principles into action. Consequently, designing solutions in the form of tools and guidelines to tackle the challenges faced by them, by working in close collaboration with practitioners would be advantageous.
    
    \item[\faThumbsUp] Investigating the users' view of ethics in AI, for example, through a similar grounded theory literature review (GTLR) approach as applied in this review to address the practitioners' view because to the best of our knowledge, this is the first grounded theory literature review (GTLR) in Software Engineering.
    \item[\faThumbsUp] Understanding the interplay between the role of practitioners and users in implementing ethics in the development and use of AI systems as one of the findings of our study shows that AI practitioners who participated in the included primary studies were less concerned about \emph{user-related} aspects when it comes to developing ethical AI systems, including human limitations, biases, and strengths.
  \end{itemize}

\section{Methodological Lessons Learned} \label{lessons}
We followed \citet{wolfswinkel2013using} guidelines to conduct our grounded theory literature review (GTLR) as it is an overarching review framework that helped us frame the review process. A grounded theory literature review (GTLR) is suitable for exploring new and emerging research areas deeply, building theories, and making practical recommendations. The process involves an iterative approach to finding relevant papers to the research topic. As per \citet{wolfswinkel2013using}, you refine the sample based on the title and abstract after removing duplicates. However, the guidelines don't provide clear steps if the return on investment is low. 
As mentioned in Section 3.3, we read the title and abstract of the first few samples (we read 200 papers) in three databases, including ACM DL, IEEEX, and SpringerLink, and all 83 papers in Wiley OL to gauge how many papers we might get. Unfortunately, this method proved inefficient, requiring full-text scans to judge relevance to our research topic. Despite considerable effort, the return on investment was minimal, with only one or two relevant papers found that included AI practitioners' views on ethics in AI for every hundred papers. This experience taught us that for a very new research topic with highly specific inclusion and exclusion criteria, it is not worth going through the titles and abstracts of all the papers in the initial search due to the expected low return on investment.

From our initial search, we found only 13 papers. Since, \citet{wolfswinkel2013using} welcome adaptations to their framework by acknowledging that ``... one size does not fit all, and there should be no hesitation whatsoever to deviate from our proposed steps, as long as such variation is well motivated", we conducted forward and backward snowballing on those 13 articles. During the snowballing process, we had to modify our \emph{seed review protocol} to find relevant papers that had information on AI practitioners’ views on ethics in AI. This significantly helped us find more relevant articles—25 more, to be precise. We discovered that employing the forward and backward snowballing method and relaxing the review protocol after identifying seed papers is a more effective way to find relevant literature, as it worked well for our research. While \citet{wolfswinkel2013using} guidelines don't explicitly mention adjusting the review protocol, they are open to adaptations. In our study, we embraced this flexibility and made modifications that proved successful for us.

\section{Conclusion} \label{conclusion}
AI systems are as ethical as the humans developing them. It is critical to understand how the humans in the trenches, the AI practitioners, view the topic of ethics in AI if we are to a lay firm theoretical foundation for future work in this area. With this in mind, we formulated the research question: \textit{What do we know from the literature about the AI practitioners’ views and experiences of ethics in AI?} To address this, we conducted a grounded theory literature review (GTLR) introduced by \cite{wolfswinkel2013using}, applying the concrete steps of socio-technical grounded theory (STGT) for data analysis and developed a \textit{taxonomy} \citep{hoda2021socio}, based on 38 primary empirical studies. Since there were not many empirical studies focusing on this niche topic exclusively, a grounded theory-based iterative and responsive review approach worked well to identify and extract relevant content from across multiple studies (that mainly focused on other related topics). The application of socio-technical grounded theory (STGT) for data analysis procedures such as open coding, constant comparison, memoing, targeted coding, and theoretical structuring enabled rigorous analysis and taxonomy development. We identified five categories of \textit{practitioner awareness}, \textit{practitioner perception}, \textit{practitioner need}, \textit{practitioner challenge}, and \textit{practitioner approach}, including the underlying concepts and codes giving rise to these categories. Taken together, and applying theoretical structuring, we developed a \textit{taxonomy of ethics in AI from practitioners' viewpoints} to guide AI practitioners, researchers, and educators in identifying and understanding the different aspects of AI ethics to consider and manage. The taxonomy serves as a research agenda for the community, where future work can focus on investigating and explaining each of the individual phenomena of practitioner awareness, perception, challenge, need, and approach in-depth. Future empirical studies can focus on improving the understanding and implementation of ethics in AI and recommend practical approaches to minimise ethical issues such as mitigating potential biases in AI development through frameworks and tools development.

\appendices
\section{List of Included Studies} \label{Appendix A}
\begin{footnotesize}

\begin{enumerate}[{G1}]
    \item Vakkuri V, Kemell K-K, Kultanen J, Siponen M, Abrahamsson P (2019) Ethically aligned design of autonomous systems: Industry viewpoint and an empirical study. arXiv preprint arXiv:1906.07946
    \item Holstein K, Wortman V J, Daume III H, Dudik M, Wallach H (2019) Improving fairness in machine learning systems: What do industry practitioners need? In Proceedings of the 2019 CHI Conference on Human Factors in Computing Systems, pp 1–16, DOI: https://doi.org/10.1145/3290605.3300830
    \item Veale M, VanKleek M, Binns R (2018) Fairness and accountability design needs for algorithmic support in high-stakes public sector decision-making. In Proceedings of the 2018 CHI Conference on Human Factors in Computing Systems, pp 1–14, DOI: https://doi.org/10.1145/3173574.3174014
    \item Vakkuri V, Kemell K-K, Kultanen J, Abrahamsson P (2020) The current state of industrial practice in artificial intelligence ethics. IEEE Software 37(4), pp 50–57, DOI:10.1109/M\\S.2020.2985621
    \item Vakkuri V, Kemell K-K, Abrahamsson P (2019) Implementing ethics in AI: Initial results of an industrial multiple case study. In International Conference on Product-Focused Software Process Improvement, pp 331–338, DOI://doi.org/10.1007/9 78-3-030-35333-9-24
    \item Ibanez J C, Olmeda M V (2021) Operationalising AI ethics: How are companies bridging the gap between practice and principles? An exploratory study. AI \& Society 37, pp 1–25, DOI: https://doi.org/10.1007/s00146-021-01267-0
    \item Orr W, Davis J L (2020) Attributions of ethical responsibility by artificial intelligence practitioners. Information, Communication \& Society 23(5), pp 719–735, DOI: 10.1080/1369\\118X.2020.1713842
    \item Rothenberger L, Fabian B, Arunov E (2019) Relevance of ethical guidelines for artificial intelligence– A survey and evaluation. In Proceedings of the 27th European Conference on Information Systems, Stockholm \& Uppsala, Sweden, DOI: https://aisel.aisnet.org/ec\\is2019\_rip/26
    \item Vakkuri V, Kemell K-K, Abrahamsson P (2019) Ethically aligned design: An empirical evaluation of the Resolvedd-strategy in software and systems development context. In 45th Euromicro Conference on Software Engineering and Advanced Applications, pp 46–50, DOI: 10.1109/SEAA.2019.00015
   \item Kelley S (2021) Employee perceptions of effective AI principle adoption. AI Principle Adoption \& Implementation, ResearchGate preprint 
   \item Madaio M, Stark L, Vaughan J W, Wallach H (2020) Co-designing checklists to understand organizational challenges and opportunities around fairness in AI. In Proceedings of the 2020 CHI Conference on Human Factors in Computing Systems, pp 1-14, DOI: http://dx.doi.org/ 10.1145/3313831.3376445
   \item Addis C, Kutar M (2019) AI management an exploratory survey of the influence of GDPR and FAT principles. In 2019 IEEE SmartWorld, Ubiquitous Intelligence \& Computing, Advanced \& Trusted Computing, Scalable Computing \& Communications, Cloud \& Big Data Computing, Internet of People and Smart City Innovation, pp 342-347, DOI: 10.1109/SmartWorld-UIC-ATC-SCALCOM-IOP-SCI.2019.00102
   \item Baker-Brunnbauer J (2021) Management perspective of ethics in artificial intelligence. AI and Ethics 1(2): 173-181, DOI: https://doi.org/10.1007/s43681-020-00022-3
   \item Frick N R, Brunker F, Ross B, Stieglitz S (2020) Design requirements for AI-based services enriching legacy information systems in enterprises: A managerial perspective. In Proceedings of 31st Australasian Conference on Information Systems (ACIS), pp 1--4
   \item Seah J, Findlay M (2021) Communicating ethics across the AI ecosystem. SMU Centre for AI \& Data Governance Research Paper
   \item Govia L (2020) Coproduction, ethics and artificial intelligence: A perspective from cultural anthropology. Journal of Digital Social Research 2(3): 42–64, DOI: https://doi.org/\\10.33621/jdsr.v2i3.53
   \item Mark R, Anya G (2019) Ethics of using smart city AI and big data: The case of four large European cities. The ORBIT Journal 2(2): 1–36, DOI: https://doi.org/10.29297/orbit.v2\\i2.110
   \item Stahl B C, Antoniou J, Ryan M, Macnish K, Jiya T (2021) Organisational responses to the ethical issues of artificial intelligence. AI \& Society, 37: 1–15, DOI: https://doi.org/10\\.1007/s00146-021-01148-6
   \item Lu Q, Zhu L, Xu X, Whittle J, Douglas D, Sanderson C (2022) Software engineering for responsible AI: An empirical study and operationalised patterns. In Proceedings of the 44th International Conference on Software Engineering: Software Engineering in Practice, pp 241--242, DOI: https://doi.org/10.1145/3510457.3513063
   \item Kessing M (2021) Fairness in AI: Discussion of a unified approach to ensure responsible AI development. Master dissertation, KTH Royal Institute of Technology
   \item Karakash T (2021) The double-edged razor of machine learning algorithms in marketing: benefits vs. ethical concerns. Dissertation, University of Twente
   \item Sun T Q, Medaglia R (2019) Mapping the challenges of artificial intelligence in the public sector: Evidence from public healthcare. Government Information Quarterly 36(2), pp 368–383, DOI: https://doi.org/10.1016/j.giq.2018.09.008
   \item Stahl B C (2021) Artificial intelligence for human flourishing–beyond principles for machine learning. Journal of Business Research 124, pp 374– 388, DOI: https://doi.org/10.1016\\/j.jbusres.2020.11.030
   \item Christodoulou E, Iordanou K (2021) Democracy under attack: Challenges of addressing ethical issues of AI and big data for more democratic digital media and societies. Frontiers in Political Science 3: 71, DOI: 10.3389/fpos.2021.682945
   \item Chivukula S S, Hasib A, Li Z, Chen J, Gray C M (2021) Identity claims that underlie ethical awareness and action. In Proceedings of the 2021 CHI Conference on Human Factors in Computing Systems, pp 1-13, DOI: 10.1145/3411764.3445375
   \item Sanderson C, Douglas D, Lu Q, Schleiger E, Whittle J, Lacey J, Newnham G, Hajkowicz S, Robinson C, Hansen D (2021) AI ethics principles in practice: Perspectives of designers and developers. arXiv preprint arXiv:2112.07467
   \item Ryan M, Antoniou J, Brooks L, Jiya T, Macnish K, Stahl B (2021) Research and practice of AI ethics: A case study approach juxtaposing academic discourse with organisational reality. Science and Engineering Ethics 27(2), pp 1–29, DOI: https://doi.org/10.1007/s11948\\-021-00293-x
   \item Chivukula S S, Watkins C R, Manocha R, Chen J, Gray C M (2020) Dimensions of UX practice that shape ethical awareness. In Proceedings of the 2020 CHI Conference on Human Factors in Computing Systems, pp 1–13, DOI: http://dx.doi.org/10.1145/3313831.3\\376459
   \item Morley J, Kinsey L, Elhalal  A, Garcia F, Ziosi M, Floridi L (2021) Operationalising AI ethics: Barriers, enablers and next steps. AI \& Society, pp 1-13, DOI: https://doi.org/10.1\\007/s00146-021-01308-8
   \item Slota S C, Fleischmann K R, Greenberg S, Verma N, Cummings B, Li L, Shenefiel C (2021) Many hands make many fingers to point: Challenges in creating accountable AI. AI \& Society, pp 1-13, DOI: https://doi.org/10.1007/s00146-021-01302-0
\item Vakkuri V, Kemell K K, Tolvanen J, Jantunen M, Halme E, Abrahamsson P (2022) How do software companies deal with artificial intelligence ethics? A gap analysis. In Proceedings of the International Conference on Evaluation and Assessment in Software Engineering, pp 100-109, DOI: https://doi.org/10.1145/3530019.3530030
\item Karen B (2022) Designing up with value-sensitive design: Building a field guide for ethical ML development. In 2022 ACM Conference on Fairness, Accountability, and Transparency, pp 2069-2083, DOI: https://doi.org/10.1145/3531146.3534626  
\item Madaio M, Egede L, Subramonyam H, Vaughan J W, Wallach H (2022) Assessing the fairness of AI systems: AI practitioners' processes, challenges, and needs for support. In Proceedings of the ACM on Human-Computer Interaction 6(CSCW1), pp 1-26, DOI: https://doi.org/10.1145/3512899
\item Widder D G, Nafus D (2023) Dislocated accountabilities in the ``AI supply chain”: Modularity and developers' notions of responsibility. Big Data \& Society 10(1), pp 20539517231177620, DOI: https://doi.org/10.1177/20539517231177620
\item Deng W H, Nagireddy M, Seng M A L, Singh J, Wu Z S, Holstein K, Zhu H (2022) Exploring how machine learning practitioners (try to) use fairness toolkits. arXiv preprint arXiv:2205.06922
\item Deng W H, Guo B B, DeVrio A, Shen H, Eslami M, Holstein K (2022) Understanding practices, challenges, and opportunities for user-driven algorithm auditing in industry practice. arXiv preprint arXiv:2210.03709 
\item Bogdana R, Yang J, Cramer H, Chowdhury R (2021) Where responsible AI meets reality: Practitioner perspectives on enablers for shifting organizational practices. In Proceedings of the ACM on Human-Computer Interaction 5(CSCW1): pp 1-23, DOI: https://do\\i.org/10.1145/3449081
\item Jiyoo C, Custis C (2022) Understanding implementation challenges in machine learning documentation. Equity and Access in Algorithms, Mechanisms, and Optimization, pp 1-8, DOI: https://doi.org/10.1145/3551624.3555301 
\end{enumerate}
 \end{footnotesize}

\section{Documentation of the Search Process} \label{Appendix B} 
\begin{figure}[ht]
    \centering
    \includegraphics [scale=0.33]{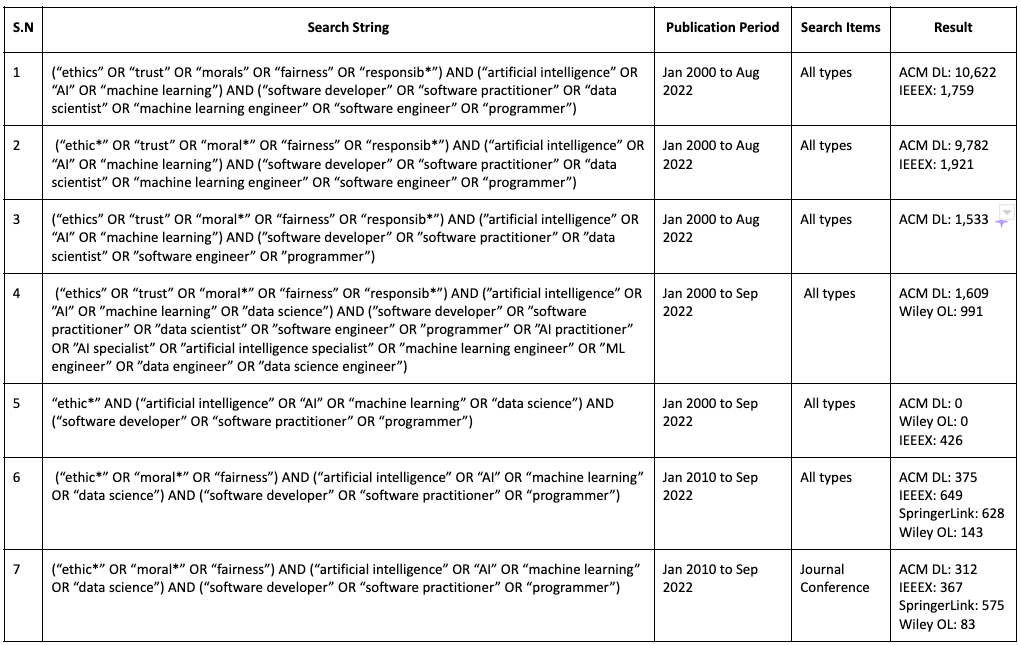} 
    \caption{Documentation of the Search Process} 
\end{figure}

\pagebreak
\section{Glossary of Terms} \label{Appendix C}
\begin{footnotesize}
In this section, we provide definitions for certain terms used in the manuscript. The definitions referenced are directly sourced, while those without citations are developed by the authors. 

\begin{itemize}
\item \textbf{Ethics}: The moral principles that govern the behaviors or activities of
a person or a group of people \citep{nalini2020hitchhiker}.
    \item \textbf{AI Ethics}: The principles of developing AI to interact with other AIs and humans ethically and function ethically in society \citep{siau2020artificial}. 
    \item \textbf{AI Practitioner}: The term ‘practitioners’ in our study includes AI developers, AI engineers, AI specialists, and AI experts. The terms ‘AI practitioners’ and ‘practitioners’ are used interchangeably throughout our study.
    \item \textbf{Fairness}: AI systems should be inclusive and accessible, and should not involve or result in unfair discrimination against individuals, communities, or groups \citep{Australia}.
    \item \textbf{Accountability}: People responsible for the different phases of the AI system lifecycle should be identifiable and accountable for the outcomes of the AI systems, and human oversight of AI systems should be enabled \citep{Australia}.
    \item \textbf{Transparency and explainability:} There should be transparency and responsible disclosure so people can understand when they are being significantly impacted by AI and can find out when an AI system is engaging with them \citep{Australia}.
\item \textbf{Privacy protection and security}: AI systems should respect and uphold privacy rights and data protection, and ensure the security of data \citep{Australia}.
\end{itemize}
\end{footnotesize}

\begin{acknowledgements}
Aastha Pant is supported by the Faculty of IT Ph.D. scholarship from Monash University. C. Tantithamthavorn is partially supported by the Australian Research Council's Discovery Early Career Researcher Award (DECRA) funding scheme (DE200100941). Also, the authors would like to thank Prof. John Grundy for his constructive feedback on the paper. 
\end{acknowledgements}

\section* {Data Availability Statement}
All data generated or analysed during this study are included in this published article (and its supplementary information files).

\section*{Declarations}
\textbf{Funding and/or Conflicts of interests/Competing interests}
The authors declare that they have no conflict of interest.

\bibliographystyle{spbasic} 
\bibliography{bibfile.bib}  
\begin{biography}{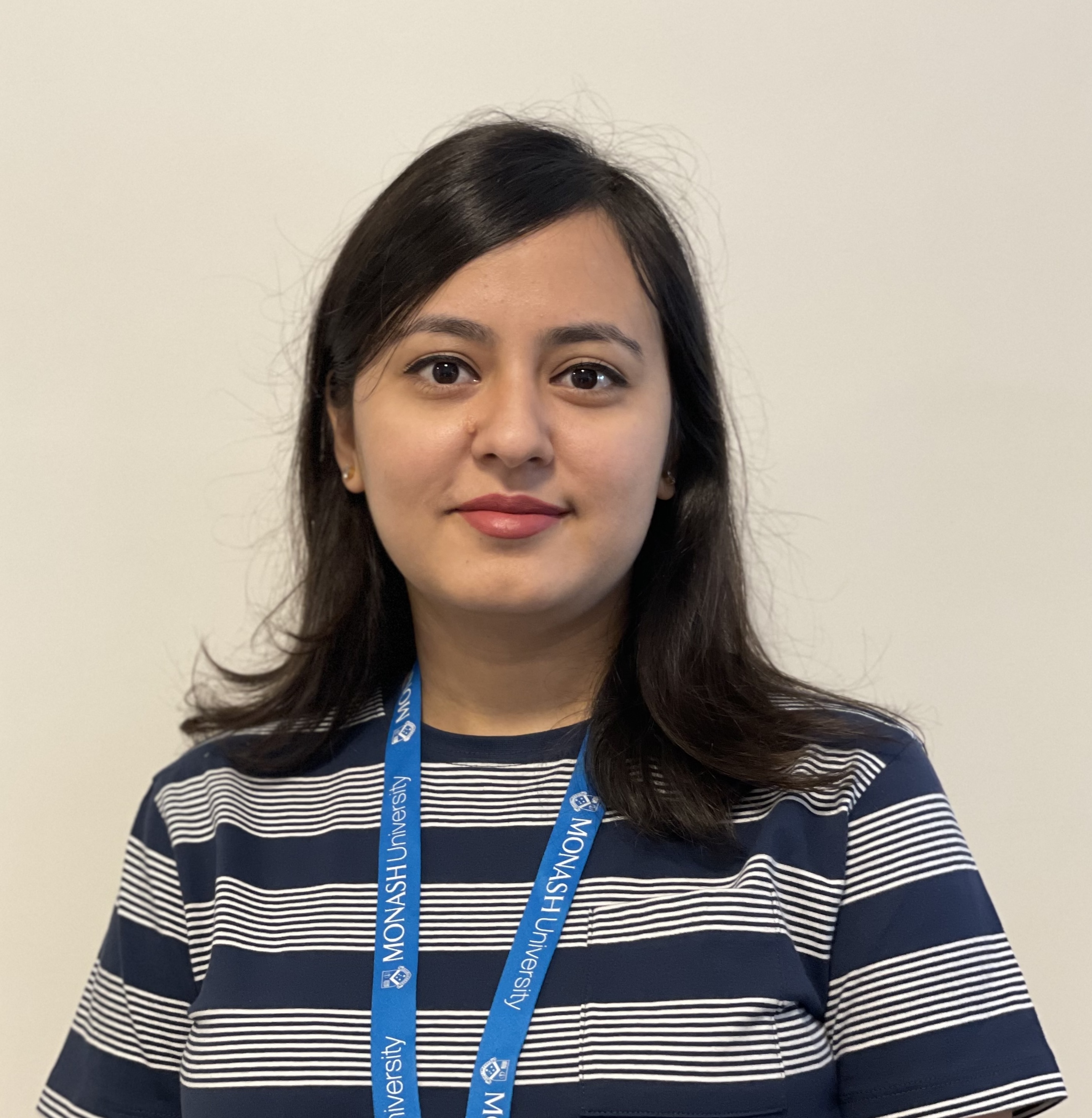} \textbf{Aastha Pant} is a Ph.D. candidate at Monash University in Melbourne, Australia. She is also currently working as a research assistant and teaching associate at Monash University. She holds a Master in Research from the University of Southern Queensland in Toowoomba, Australia. Before embarking on her Ph.D. journey, she was in academia as a teaching assistant. Her research interests encompass a broad spectrum, with a focus on areas such as ethics in artificial intelligence, and socio-technical aspects of software engineering. More details of her research can be found at, \protect{\url{https://www.researchgate.net/profile/Aastha-Pant-3}} Contact her at: aastha.pant@monash.edu.
\end{biography}
 \vspace{1cm}
\begin{biography}{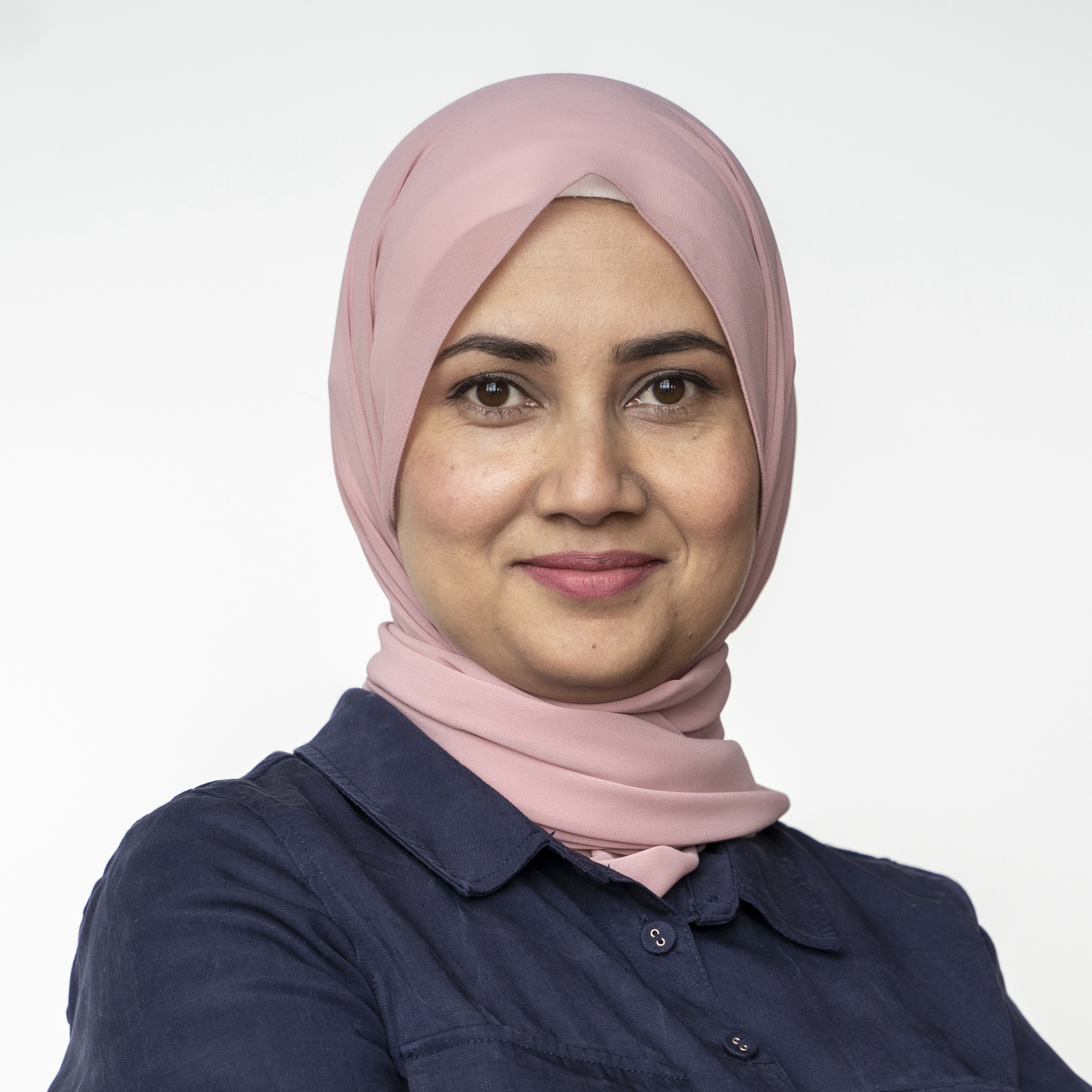}\textbf{Rashina Hoda} is an Associate Professor of Software Engineering at Monash University, Melbourne, specialising in the human and socio-technical aspects of software engineering and artificial intelligence including agile methods, ethics in AI, and human values. Rashina has introduced Socio-Technical Grounded Theory (STGT) as a modern variant of traditional Grounded Theory to Software Engineering. She serves as an Associate Editor of the IEEE Transactions on Software Engineering and as ICSE 2024 Workshops co-chair. Previously, she served as co-chair for ICSE-SEIS 2023, PC co-chair for CHASE 2021, Associate Editor of JSS, and on the advisory board of IEEE Software. For more: \protect{\url{www.rashina.com}}.
\end{biography}
  \vspace{1cm}
\begin{biography}{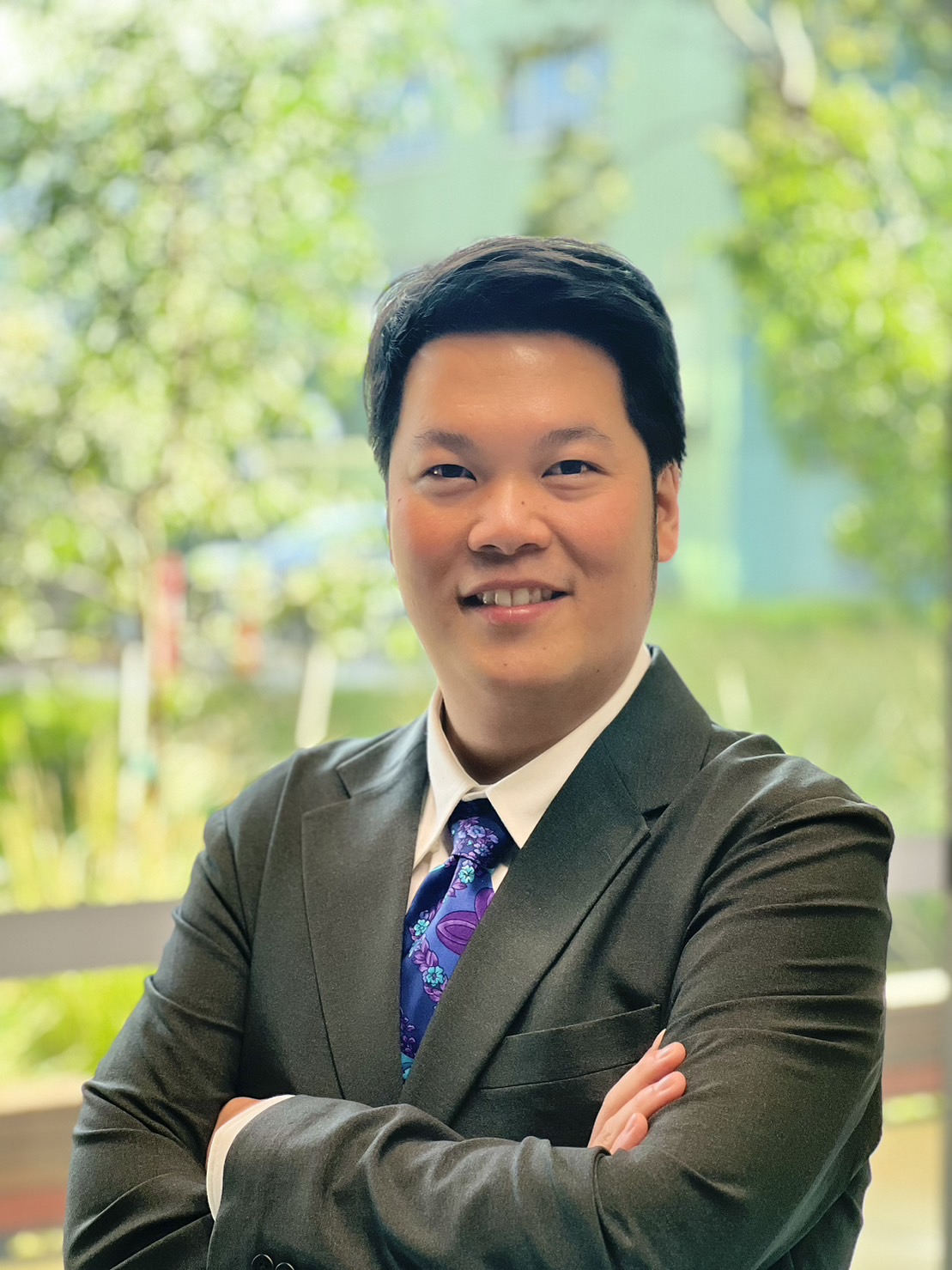}\textbf{Chakkrit (Kla) Tantithamthavorn} is a Senior Lecturer in Software Engineering at the Faculty of Information Technology, Monash University, Australia. He is pioneering an emerging research area of explainable AI for software engineering, inventing many AI-based technologies to improve developers’ productivity and make software systems more reliable and more secure while being explainable to practitioners. He has made several major advances in explainable AI for software engineering and published the first online book on explainable AI for software engineering (http:// xai4se.github.io), attracting 20,000+ page-views from 83 countries worldwide and receiving positive responses from the SE community. His publications, books, and tutorials have informed many other studies and educated the SE community on the importance of explainability and its applications to software engineering. More about him is available at \protect{\url{http://chakkrit.com}}.
\end{biography}
 \vspace{1cm}
\begin{biography}{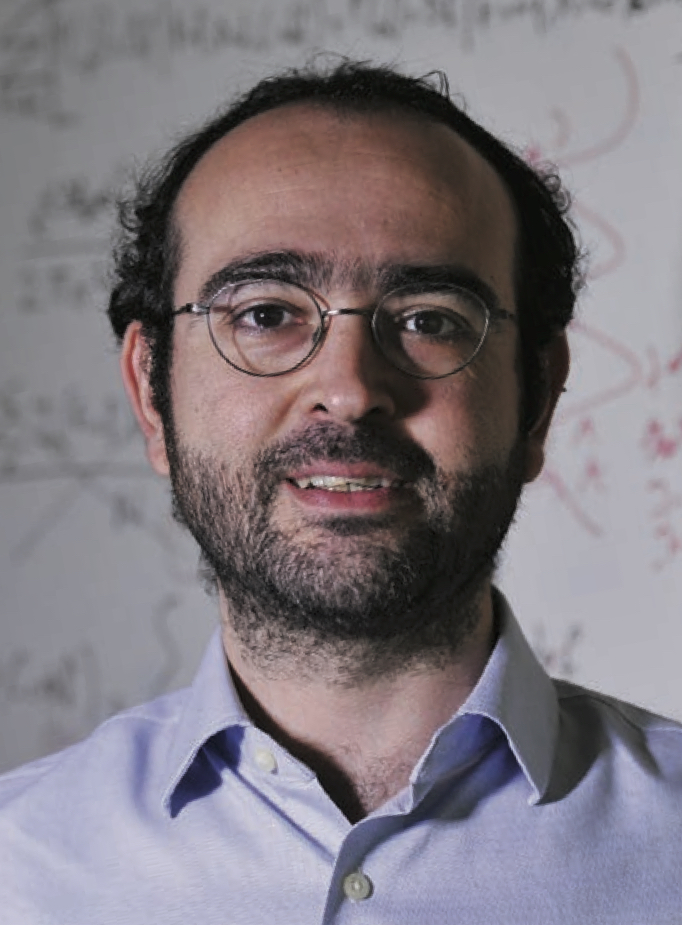}\textbf{Burak Turhan} Ph.D. (Bo\~{g}azi\c{c}i University), is a Professor in the M3S Research Unit at the University of Oulu, Finland, and an Adjunct Professor (Research) at Monash University, Australia. His research focuses on empirical software engineering, artificial intelligence, quality assurance and testing, human factors, and (agile) development processes. Dr. Turhan has published over 120 articles in international journals and conferences, received several best paper awards, and secured funding for several large-scale research projects. He has served on the editorial boards of several software engineering journals (EMSE, ACM TOSEM, JSS, ASE, IST, SQJ), as (co-)chair for PROMISE'13, ESEM'17, PROFES'17, and EASE'23, and as a steering committee member for PROMISE and ESEM. He is a member of the ACM, ACM SIGSOFT, IEEE, and IEEE Computer Society. For more information, please visit: \protect{\url{https://turhanb.net}}.
\end{biography}
 
\end{document}